\documentclass[12pt,preprint]{aastex}

\newcommand{\snr}{G292.0$+$1.8}

\slugcomment{Submitted: \today}
\shorttitle{AKARI Infrared Observations of G292.0$+$1.8}
\shortauthors{Lee et al.}

\begin{document}
\def\kms{km~s$^{-1}$}
\def\vecJ{{\bf J}}
\def\mum{$\mu$m}
\def\neontwo{[\ion{Ne}{2}]}
\def\neonthree{[\ion{Ne}{3}]}
\def\othree{[\ion{O}{3}]}
\def\simlt{\lower.5ex\hbox{$\; \buildrel < \over \sim \;$}}
\def\simgt{\lower.5ex\hbox{$\; \buildrel > \over \sim \;$}}

\title{AKARI Infrared Observations of the Supernova Remnant 
G292.0+1.8:
Unveiling Circumstellar Medium and Supernova Ejecta}

\author{Ho-Gyu Lee\altaffilmark{1, 2}, 
Bon-Chul Koo\altaffilmark{2},
Dae-Sik Moon\altaffilmark{3}, 
Itsuki Sakon\altaffilmark{1}, 
Takashi Onaka\altaffilmark{1}, 
Woong-Seob Jeong\altaffilmark{4},
Hidehiro Kaneda\altaffilmark{5},
Takaya Nozawa\altaffilmark{6}, 
and Takashi Kozasa\altaffilmark{7}
}

\altaffiltext{1}{Department of Astronomy, Graduate School of Science,
The University of Tokyo, Bunkyo-ku, Tokyo 113-0033, Japan; 
hglee@astron.s.u-tokyo.ac.jp
isakon@astron.s.u-tokyo.ac.jp,
onaka@astron.s.u-tokyo.ac.jp}
\altaffiltext{2}{Department of Physics and Astronomy, 
Seoul National University, Seoul, 151-742, Korea; 
koo@astrohi.snu.ac.kr}
\altaffiltext{3}{Department of Astronomy and Astrophysics, 
University of Toronto, Toronto, ON M5S 3H4, Canada; 
moon@astro.utoronto.ca}
\altaffiltext{4}{Korea Astronomy and Space Science Institute, 
61-1, Whaam-dong, Yuseong-gu, Deajeon 305-348, Korea;
jeongws@kasi.re.kr}
\altaffiltext{5}{Graduate School of Science, Nagoya University, 
Chikusa-ku, Nagoya 464-8602, Japan;
kaneda@u.phys.nagoya-u.ac.jp}
\altaffiltext{6}{Institute for the Physics and Mathematics of the
Universe, University of Tokyo, Kashiwa, Chiba 277-8568, Japan;
tnozawa@mail.sci.hokudai.ac.jp}
\altaffiltext{7}{Department of Cosmosciences, Graduate School
of Science, Hokkaido University, Sapporo 060-0810, Japan;
kozasa@mail.sci.hokudai.ac.jp}

\begin{abstract}

We present the results of $AKARI$ 
observations of the O-rich supernova remnant G292.0$+$1.8
using six IRC and four FIS bands 
covering 2.7--26.5 $\mu$m and 50--180 $\mu$m, respectively.
The $AKARI$ images show two prominent structures; 
a bright equatorial ring structure along the east-west direction 
and an outer elliptical shell structure surrounding the remnant.
The equatorial ring structure is clumpy 
and incomplete with its western end opened. 
The outer shell is almost complete and slightly squeezed 
along the north-south direction. 
The central position of the outer shell is $\sim$ 1$'$ northwest from
the embedded pulsar 
and coincides with the center of the equatorial ring structure.
In the northen and southwestern regions, there is also faint emission
with a sharp boundary beyond the bright shell structure. 
The equatorial ring and the elliptical shell structures
were partly visible in optical and/or X-rays, 
but they are much more clearly revealed in our $AKARI$ images. 
There is no evident difference in infrared colors of the two prominent structures, 
which is consistent with the previous proposition that both structures 
are of circumstellar origin. 
However, we have detected faint infrared emission of a considerably high 
15 to 24 $\mu$m ratio associated with the supernova ejecta in
the southeastern and northwestern areas.
Our IRC spectra show that the high ratio is 
at least partly due to the emission lines from Ne ions in the supernova ejecta material.
In addition we detect a narrow, elongated
feature outside the SNR shell.
We derive the physical parameters of the infrared-emitting dust grains 
in the shocked circumstellar medium
and compare the result with model calculations 
of dust destruction by a SN shock.
The $AKARI$ results suggest that
the progenitor
was at the center of the infrared circumstellar shell in red supergiant
stage and that 
the observed asymmetry in the SN ejecta
could be a result of either a dense
circumstellar medium in the equatorial plane and/or an asymmetric
explosion.
\end{abstract}

\keywords{ISM: individual(\objectname{G292.0+1.8}) --- 
infrared: ISM ---
ISM: dust ---
shock waves ---
supernova remnants
}

\section{Introduction}

Young core-collapse supernova remnants (SNRs) in the Galaxy provide
an unique opportunity to study fine details of ejecta from supernova (SN) explosions
as well as those of the circumstellar medium (CSM) produced over the final evolutionary
stages of massive stars.
An obvious group of such young core-collapse SNRs is O-rich SNRs 
that show 
optical spectra featuring strong O and Ne lines, with lines from lighter elements 
(e.g., He) either weak or absent.
Since the ejecta from a progenitor and the swept-up CSM in the O-rich SNRs
are not completely mixed together yet, observation of the O-rich SNRs is promising to study
the ejecta and CSM related to the explosion of a progenitor and its late-stage evolution.

G292.0$+$1.8 (MSH11--54), 
together with Cassiopeia~A and Puppis~A, forms a rare group of O-rich SNRs in the Galaxy.
The O-rich nature of \snr\ was discovered by the optical detection of fast-moving 
O-rich and Ne-rich ejecta \citep{gos79, mur79}.
Recent wide-field observations of the optical \othree\ line in \snr\ have revealed that 
the ejecta velocities range from 
$-$1400~\kms\ to $+$1700~\kms\
and also that distant ejecta knots are located primarily 
along the north-south direction \citep{gha05, win06}.
The dynamical center of G292.0$+$1.8, 
based on the kinematics of the \othree\ line emission from the ejecta,
is roughly coincident with the geometrical center of the radio emission, 
but shows an apparent offset from its pulsar position 
discovered at the southeast \citep[][see below]{win08}. 
\cite{gha09} reported that the mid-infrared (MIR) spectrum of the ejecta is dominated by
ionic lines and a broad bump around 17~$\mu$m using 
the $Spitzer$ Space Telescope ($Spitzer$).
They proposed that
the latter is produced either by Polycyclic Aromatic Hydrocarbons (PAHs)
along the line of sight 
or newly-formed dust within the ejecta.
Besides there is also an equatorial MIR continuum component from the dust 
associated with the shocked CSM partly overlapping with the ejecta emission.
A previous infrared (IR) study based on the data obtained with 
Infrared Astronomical Satellite ($IRAS$), on the other hand, 
identified enhanced far-IR (FIR) emission in the southwestern part of the SNR, 
suggesting that the source has encountered adjacent clouds \citep{bra86, par07}.

In radio, \snr\ mainly consists of bright emission from a central pulsar
wind nebula (PWN) of $\sim$~4\arcmin\ (in diameter) and relatively fainter outer plateau emission 
of $\sim$~9\arcmin\ \citep{loc77, bra86, gae03}.
The plateau emission declines sharply outward, and there is no apparent surrounding radio shell structure.
The pulsar PSR J1124--5916, which is located $\sim$46\arcsec\ from 
the dynamical center of the SNR in the southeast direction,
was discovered by radio timing \citep{cam02}, 
confirming the PWN nature of the central emission of the SNR.
The pulsar and PWN were also identified in X-ray emission \citep{hug03},
where the pulsar has a nearby jet and a torus of $\sim$~5\arcsec\ scale \citep{par07}.
The jet axis is slightly tilted to the northeast direction \citep{par07}. 
There are two clear differences between the soft and hard X-ray emission of \snr.
First, while the former is produced by shocked gas of 
normal composition distributed as equatorial filaments,
indicating it is likely from CSM,
the latter is dominated by bright metallic lines from the ejecta \citep{par02, par04}.
Next, the X-ray emission is harder in the northwestern part than the southeastern part,
implying density variation or asymmetric SN explosion \citep{par07}.
The distance to \snr, on the other hand, 
was estimated to be 6.2 kpc from H~I absorption observations \citep{gae03}.

In this paper we present an extensive IR study of G292.0$+$1.8 
using multi-band imaging and spectroscopic data obtained with the $AKARI$ 
space telescope \citep{mur07}.
We describe the details of our $AKARI$ observations 
and present the results of the observations in \S2 and 3, respectively.
In \S4 we discuss IR emission in \snr\ from circumstellar dust 
with a comparison to model calculations of dust destruction by a SN shock.
We also discuss the IR emission from the ejecta and
we suggest that there is a negligible amount of dust associated with the ejecta.
Then, we make a comparison to the results of $Spitzer$ obserations.
We finally discuss the SN explosion in \snr\ based on our
$AKARI$ observations followed by our conclusions in \S5.

\section{Observations}

The $AKARI$ multi-band imaging observations of G292.0$+$1.8 were carried out
using its six near-IR (NIR) and MIR bands in the 2.7--26.5 $\mu$m range as well as
four FIR bands in the 50--180 $\mu$m range on 2007 January 17 and 19.
The NIR and MIR images were obtained with Infrared Camera (IRC)
that has three simultaneously operating NIR, MIR-S, and MIR-L channels.
The three channels have comparable field of views of $\sim 10'\times10'$.
The NIR and MIR-S channels share the same pointing direction, 
while that of the MIR-L channel is 25\arcmin\ away.
The IRC observations were conducted in the two-filter mode that produced 
images of two bands for each channel \citep{ona07}.
The total on-source integration times were 178~s for the NIR (N3, N4) observations
and 196~s for both the MIR-S (S7, S11) and MIR-L (L15, L24) observations.
The basic calibration and data reduction,
including dark subtraction, linearity correction, distortion correction,
flat fielding, image stacking, and absolute position determination
were performed by the standard IRC Imaging Data Reduction Pipeline 
(version 20071017)\footnote{http://www.ir.isas.jaxa.jp/ASTRO-F/Observation/DataReduction/IRC/}.
Tables~\ref{tab_obssum} and \ref{tab_obsimg} present a journal of our $AKARI$ observations, 
including the spectroscopic observations,
and the basic parameters of the $AKARI$ imaging bands, respectively.

The FIR images were obtained with Far-Infrared Surveyor (FIS)
in two round-trip scans using the cross-scan shift mode \citep{kaw07}.
The scan speed and length were 15\arcsec~s$^{-1}$ and 240\arcsec, respectively,
creating images of 40\arcmin\ $\times$ 12\arcmin\ size elongated along the scanning direction.
All the four FIS band (N60, Wide-S, Wide-L, and N160) images were obtained simultaneously 
by a single observing run (Table~\ref{tab_obssum}).
The initial data calibration and reduction such as 
glitch detection, dark subtraction, flat fielding and flux calibration
were processed with FIS Slow-Scan Toolkit 
(version 20070914)\footnote{http://www.ir.isas.jaxa.jp/ASTRO-F/Observation/DataReduction/FIS/},
and the final image construction was performed with a refined sampling mechanism.

These IRC and FIS multi-band imaging observations were followed by NIR and MIR
spectroscopic observations and L18W-band (13.9--25.6 $\mu$m, combining both L15 and L24 bands) 
imaging observations carried out in 2007 July 20--22. 
The spectroscopic observations were conducted using four (NG, SG1, SG2, and LG2) grisms.
Similar to the aforementioned imaging observations,
the NG, SG1, and SG2 mode observations were conducted simultaneously, 
while LG2 mode observations were done separately \citep{ohy07}.
Table~\ref{tab_obssp} lists the characteristics of the $AKARI$ spectroscopic observations.
Spectra from the peaks of the equatorial emission and ejecta identified
in the ratio between the L15- and L24-band images (L15/L24 hereafter) were obtained
together with that from a reference background position at the southeastern part of the source
(Table~\ref{tab_sppos}).
The data calibration and reduction were processed with IRC Spectroscopy 
Toolkit (version 20070913)\footnote{http://www.ir.isas.jaxa.jp/ASTRO-F/Observation/DataReduction/IRC/}.
The obtained flux was converted to the surface brightness based on the measured slit size.  
IRC spectroscopy is made at the slits located at the edge
of the large imaging field-of-view (FoV).
There is internal scattered light in the array of the MIR-S and
the light diffusing from the imaging FoV
affects  SG1 and SG2 spectra to some extent. This effect has
been corrected for according to the method given in \citet{sak08}.
A similar scattered light is also recognized in
LG2 spectra and has been corrected for in a similar way.
In addition there is a contribution from the second order light
from the edges of the imaging FoV in LG2 spectra,
but we expect that it is not significant in the background-subtracted spectra.

These spectroscopic observations were simultaneously accompanied by supplemental short (49~s $\times$ 3) 
L18W-band imaging observations of a field $\sim$~5\arcmin\ away from the central 
slit position in the southeast direction in order to check the pointing of the satellite. 
This supplemental, short-exposure L18W-band image of the southeast,
which fortuitously revealed interesting structures beyond the SNR boundary (see \S~3.4),
was combined with deep (442~s) L18W-band image of the source to generate a final larger-area map.
The L18W-band imaging data were processed with the same procedures that we used
for other imaging data sets as described above.

\section{Results}

\subsection{Multi-band Infrared Images of \snr}

\begin{figure}
\plotone{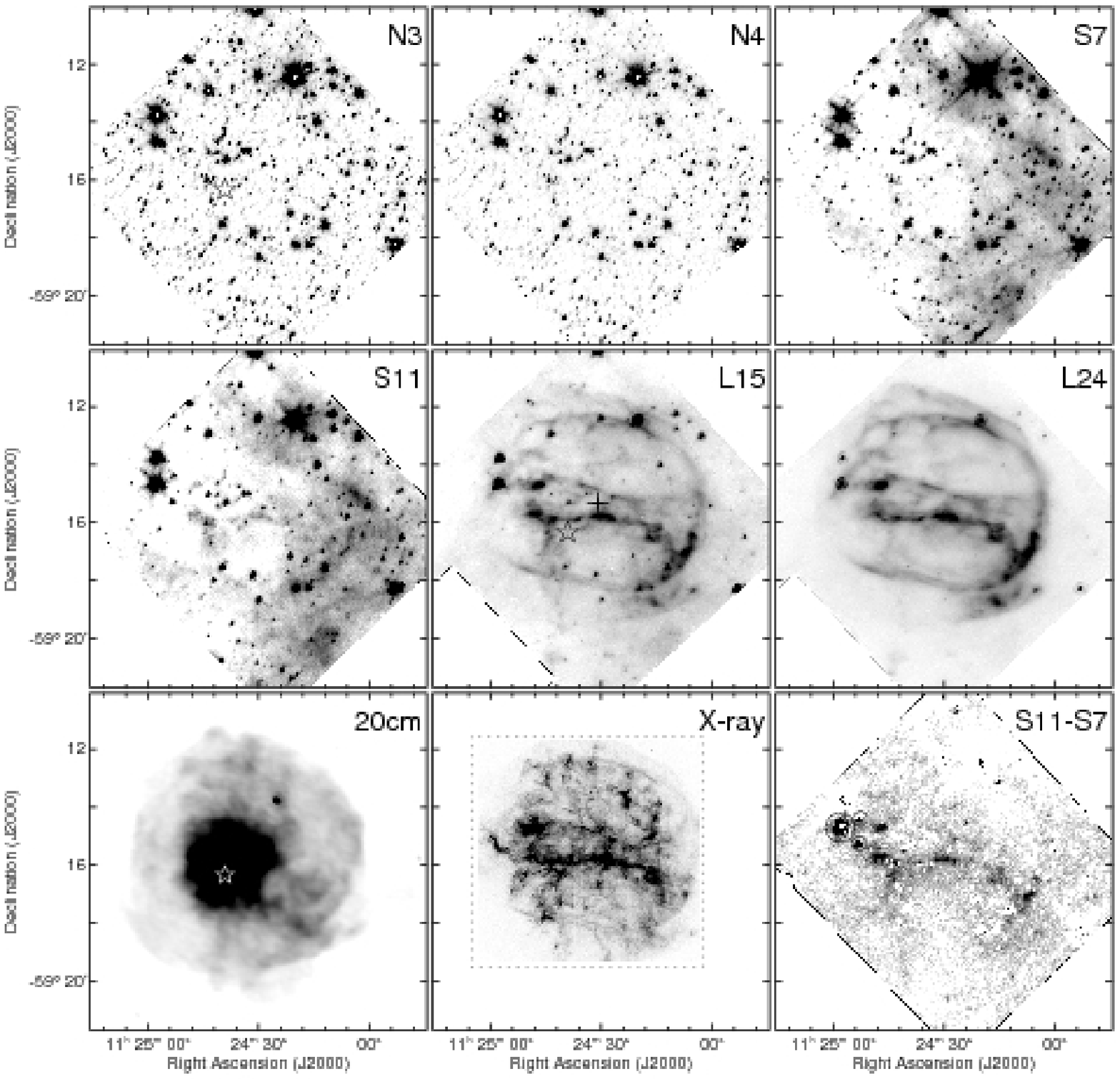}
\caption{$AKARI$ IRC images of G292.0+1.8 
with the ATCA 20 cm continuum and Chandra X-ray map for comparison
\citep{gae03, par02}. 
The background-subtracted S11-band image (S11--S7 difference image) 
is shown at the lower-right panel.
The three circles close to the eastern end of the ER in the S11--S7 difference image 
indicate stellar sources.
We mark the position of pulsar PSR J1124-5916 in N3-, L15-band \citep{hug03},
and 20 cm continuum images by the star \citep{gae03}. 
We also mark the center of OES by the plus sign in L15 image.
The dotted box in the X-ray map indicates the imaging area of Chandra.
The gray-scales of N3, N4, S7, S11, L15, L24 and S11--S7 bands are 
0.5--3.0, 0.5--3.0, 5.3--6.3, 15.8--16.7, 23--26, 29--40,
and 0--0.4 MJy sr$^{-1}$, respectively.
}
\label{fig_irc}
\end{figure}

Figure~\ref{fig_irc} presents our $AKARI$ IRC multi-band images of G292.0$+$1.8 
together with the ATCA 20~cm radio continuum image \citep{gae03},
$Chandra$ X-ray (0.3--8 keV) image \citep{par02},
and the background-subtracted $AKARI$ S11-band image (S11--S7; see below) 
for comparison.
The $AKARI$ IR emission associated with the SNR is most apparent 
in the L15- and L24-band images as two prominent features: 
first, there is a ring-like structure composed of two clumpy, narrow, and long 
filaments crossing the central part of the SNR along the east-west
direction -- we name this ``Equatorial Ring" (ER); 
secondly, the outer boundary of the SNR appears as an almost-complete
shell structure with its eastern part opened -- we name this ``Outer Elliptical Shell" (OES).

The radius of the ER is $\sim$ 3$'$ (or $\sim$ 5 pc).
Its southern filament is brighter than the northern one by a factor of 
$\sim 1.7$ on average and
has the brightest clump at 
(RA, decl.) = ($\rm 11^h24^m29^s.4$, $\rm -59^\circ15'48''$)
close to the filament center.
The southern filament also shows prominent X-ray and [\ion{O}{3}] emission 
and has been called ``equatorial bar" or ``equatorial belt" \citep{par02, gon03, gha05}.
The northern filament 
is $\sim$~1\arcmin\ away from the southern filament and 
both filaments are elongated roughly parallel to each other.

The OES shows an almost-complete ellipse of 
$\sim$ 7\arcmin\ $\times$ 6\arcmin\ ($\sim$ 12 pc $\times$ 10 pc)
with its major axis aligning roughly with the plane of the ER.
The center of the ellipse determined by elliptical fits weighted 
by the 24 $\mu$m surface brightness of the OES is 
($\rm 11^h24^m30^s.9$, $\rm -59^\circ15'21''$)
located at the middle of two filaments of the ER.
In the west the OES appears to be connected with the two filaments of the ER
and its emission is enhanced there, 
especially at the southwest near the southern filament of the ER.
In contrast the eastern part of the OES is open, 
and also the emission from the southern filament of the ER is truncated 
in the east before it reaches the boundary of the SNR.

In addition to the ER and OES, 
there appear to be at least two more clearly identifiable features in the $AKARI$ images.
First, the L15-band image shows a separate structure that extends $\sim$ 5$'$
southward from the ER on the southeastern side of the SNR. It is elongated 
vertically with the bright portion located $\sim$ 0.5$'$ below the southern 
equatorial filament.
Next, there is faint emission extending 
beyond the northern and southeastern edges of the OES.
This faint emission is contained within the outermost boundary of 
radio emission observed in G292.0$+$1.8 \citep{gae03}.

As in Figure~\ref{fig_irc}, \snr\ is not detected in the S7- and S11-band images:
while some of the S11-band emission appears to arise from locations of
strong L15- and L24-band emission,
there is almost no corresponding emission in the S7-band image. 
This suggests that the S7-band emission is dominated by background emission, 
probably emission from PAHs in the line of sight interstellar medium (ISM).
In the S11--S7 difference image (Figure~\ref{fig_irc}) we subtract out scaled S7-band 
emission from the S11-band image and confirmed that the two filaments of the ER
and southwestern part of the OES have appreciable S11-band emission.
(Note that the three bright point-like sources close to the eastern end of 
the southern filament of the ER in the S11--S7 difference image are stellar sources.)
The NIR N3- and N4-band images, on the other hand, show only stellar emission
without any apparent feature brighter than 16 $\mu$Jy at 3 $\mu$m that might 
be associated with the SNR.

Table~\ref{tab_f} presents IR measurements of G292.0$+$1.8.
In the Table, we list the fluxes and ratios 
of whole remnant area and IR-Ejecta region
(see befow for IR-Ejecta).
In addition, 
we also list the peak intensities and ratios
of the ER, the southweastern OES, and the IR-Ejecta.

\subsection{Infrared Colors and Ejecta Identification}

\begin{figure}
\plotone{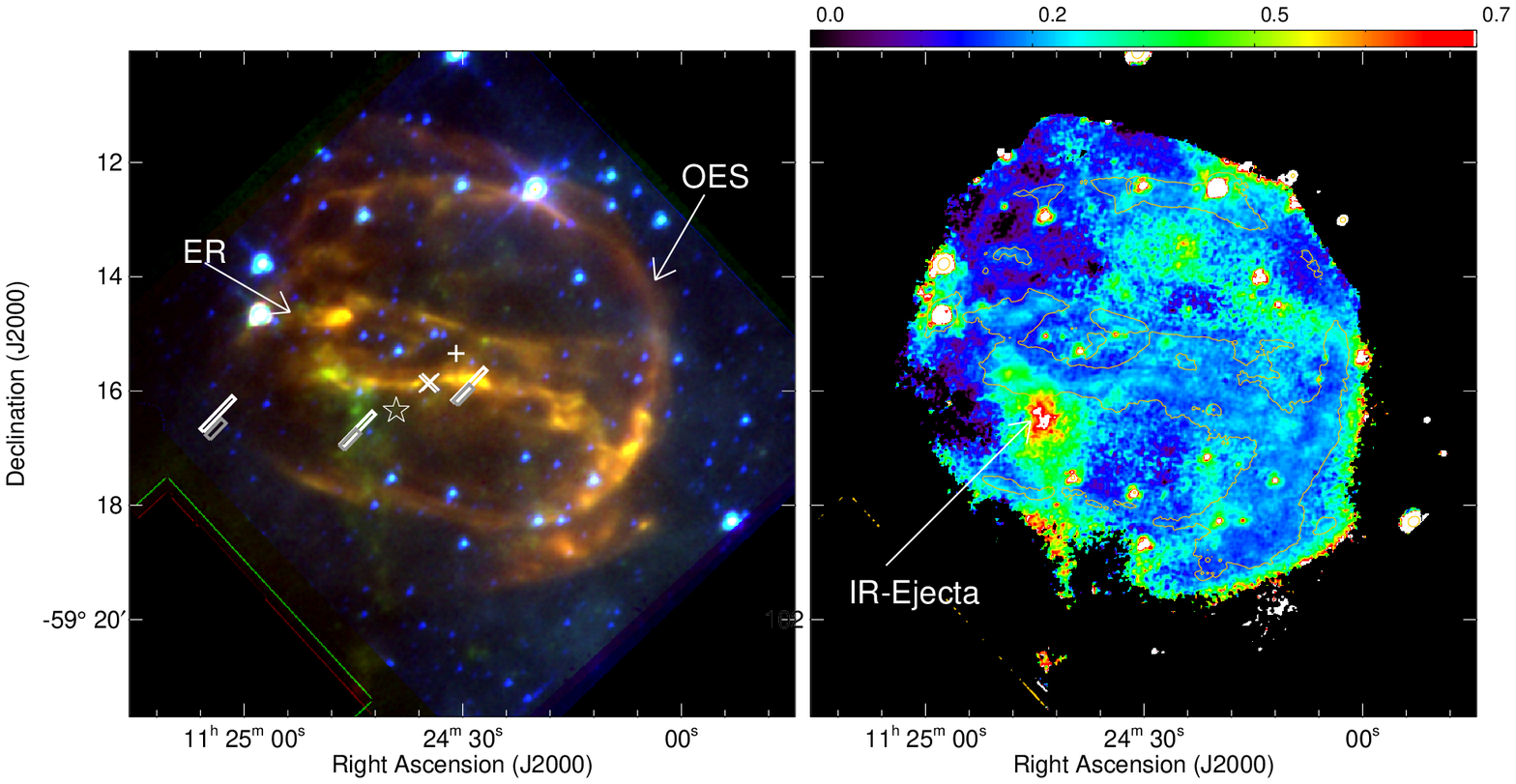}
\caption{(Left) The three-color image of G292.0$+$1.8
produced from 24 $\mu$m (R), 15 $\mu$m (G), and 7 $\mu$m (B).
The coordinates of equatorial, ejecta and background NG/SG slit positions are
indicated by white box at  
($\rm 11^h24^m29^s.1$, $\rm -59^\circ15'54''$),
($\rm 11^h24^m44^s.3$, $\rm -59^\circ16'40''$), and
($\rm 11^h25^m03^s.6$, $\rm -59^\circ16'24''$), respectively.
At the southwestern side of them, 
the three LG2 slit positions are indicated by gray box at 
($\rm 11^h24^m29^s.9$, $\rm -59^\circ16'04''$),
($\rm 11^h24^m45^s.4$, $\rm -59^\circ16'51''$), and
($\rm 11^h25^m03^s.8$, $\rm -59^\circ16'40''$), respectively.
We mark the center of the OES by plus sign,
both the dynamical center of optical ejecta knots \citep{win08}
and the center of outer radio plateau \citep{gae03} by crosses,
and the position of pulsar PSR J1124-5916 by star \citep{hug03}. 
The two cross symbols are almost coincident.
(Right) The background subtracted 15 to 24 $\mu$m ratio image.
The pixels with low 24 $\mu$m intensities are blanked out.
The bright structures in 24 $\mu$m images are shown by contour.
The prominent features such as equatorial ring (ER),
outer elliptical shell (OES), and IR-Ejecta
are also labeled.
}
\label{fig_cmap}
\end{figure}

Figure~\ref{fig_cmap} (left), which is a three-color (7, 15 and 24 $\mu$m) 
$AKARI$ MIR image of \snr, 
shows that most of the prominent features have roughly similar MIR colors.
One exception is the feature in the southeastern part of the SNR
with significant excess in the shorter wavebands --
we name this feature ``IR-Ejecta" 
because it is believed to originate in shocked SN ejecta 
as we describe below. 
It is bright near the southern equatorial filament and 
stretching directly southward
beyond the SNR boundary. 
This feature is clearly seen in Figure~\ref{fig_cmap} (right) 
which shows the L15/L24 emission ratio.
The ratio image was produced by dividing the L15-band image 
by the L24-band image after background subtraction.
The background was estimated by fitting a slanted plane to the areas 
surrounding the SNR. The resulting background planes were almost flat, 
i.e., the brightness differences over the entire images were only 
1.7 \% and 2.4 \% in L15 and L24, respectively.
We applied a Gaussian convolution to the L15-band image in order to match
the final spatial resolution to that of the L24-band image and 
masked out the pixels with small ($\le$ 30.2 MJy sr$^{-1}$) L24-band intensities.

The resulting L15/L24 color image is significantly different from the original band images.
In Figure~\ref{fig_cmap}, the L15/L24 ratio is almost uniform ($\sim$ 0.25), 
and there is no feature corresponding to the prominent ER and OES.
The most notable feature is the IR-Ejecta, 
where L15/L24 ratio raises to $\sim$~0.8,
which is in fact consistent with what we identified in the three-color image
(left panel of Figure~\ref{fig_cmap}).
It appears to be of a triangular shape with one of its vertices 
towards the direction to the SNR center.
The southern vertex passes though the southern boundary of the OES and extends over the SNR boundary,
while the northeastern vertex is located just above the ER.
The region of high color ratio of the IR-Ejecta covers a large portion of 
the southeastern area of the SNR
and positionally coincides with the O- and Ne-dominant ejecta identified 
by previous optical and X-ray observations \citep{gha05, win06, par02}.
There is also a wispy patch of emission around ($\rm 11^h24^m43^s.4$, $\rm -59^\circ20'47''$)
far ($>$ 1\arcmin) beyond the SNR boundary,
which has the color ratio similar to that of the IR-Ejecta. 
It, however, is not shown in the [\ion{O}{3}] images \citep{win06}.
Besides the IR-Ejecta in the southeastern region, the northwestern region shows extended emission
of the high color ratio, although it is not as significant as the southeastern region.
Overall the two regions of the high color ratio are roughly symmetric with respect to the center of the OES.

\begin{figure}
\plotone{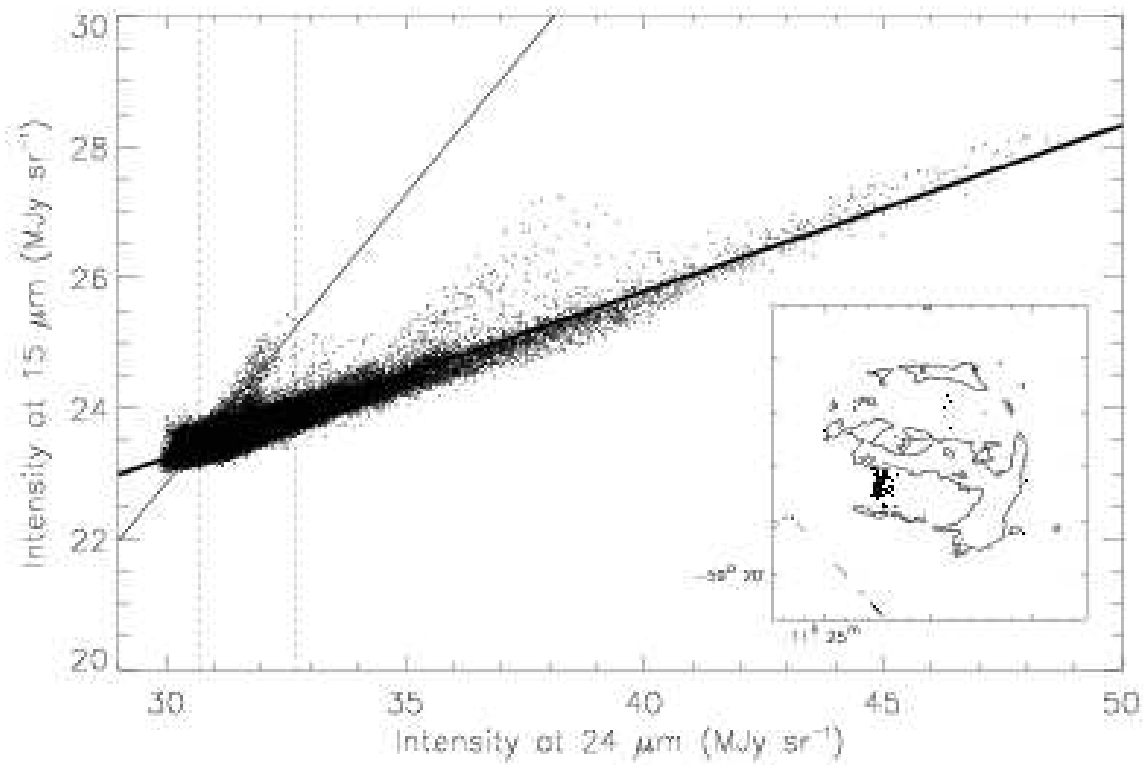}
\caption{The L15 vs L24 band intensity plot.
We mask the pixels with stellar emission using N3 band 
and the pixels with low intensity using L24 band. 
The vertical lines indicate 
the boundaries of the points with tight correlation (emission only from ejecta).
The thick line is a linear fit using all pixels in the plot.
The thin line is a linear fit using the high L15/L24 ratio region between 
two vertical lines.
The position of high L15/L24 ratio pixels are marked in the inset 
in the lower-right corner superposed on the L24 contour.
Black dots are the positions of the high L15/L24 ratio pixels 
between two vertical lines, and gray dots are those of 
high L15/L24 ratio pixels
with L24 intensity $\simgt 33$ MJy sr$^{-1}$
(ejecta emission superposed on swept-up material).
}
\label{fig_cplot}
\end{figure}

Figure~\ref{fig_cplot} compares the pixel values of the L24- and L15-band emission,
where most of the points are distributed along the thick, solid line of a slope of 0.25.
(Note that pixels of stars and pixels with small L24-band intensities 
are masked out.)
If L24- and L15-band fluxes are from the same region and if they are both
thermal dust emission, then the color ratio and the dust temperatures are related as 
\begin{equation}
\frac{I_\nu(15)}{I_\nu(24)} = 
\frac{\kappa_\nu(15) B_\nu(15,T)}{\kappa_\nu(24) B_\nu(24,T)} ~,
\end{equation}
where 
$I_\nu(\lambda)$ is the surface brightness at $\lambda$ ($\mu$m) in Jy sr$^{-1}$, 
$B_\nu(T)$ is the Planck function, 
and $\kappa_\nu(\lambda)$ is the dust opacity in cm$^2$ g$^{-1}$.  
The slope of 0.25 corresponds to the dust temperature of 126~K
for a mixture of carbonaceous and silicate interstellar grain 
of $R_{\rm V}$ = 3.1 \citep{dra03}.
In Figure~\ref{fig_cplot} there are two regions, where the data points show
a high L15/L24 ratio
compared to the thick, solid line:
first, the thin line of the slope of 0.88 represents one group whose
L24-band surface brightness is less than $\sim$~33~MJy~sr$^{-1}$;
secondly, there is another group of data points of higher L15-band emission 
whose L24-band surface brightness lies in the range of 33--40 MJy~sr$^{-1}$.
The inset in the lower-right corner of Figure~\ref{fig_cplot} shows 
that the data points with high L15/L24 ratios 
are all from the IR-Ejecta as expected. 
The ones in the second group, 
i.e., ones with high 24 $\mu$m surface brightness, 
are superposed on the ER and therefore their emission is 
partly from the swept-up CSM 
while the data points in the first group should represent 
the emission only from the ejecta. 
The slope of 0.88 corresponds 
to a dust color temperature of 240 K; however the L15-band flux is
largely from line emission at this position
so that the physical dust temperature should be lower than
this (see \S 3.4).

\subsection{Mid-Infrared Spectroscopy of Ejecta and Equatorial Peak}

\begin{figure}
\plotone{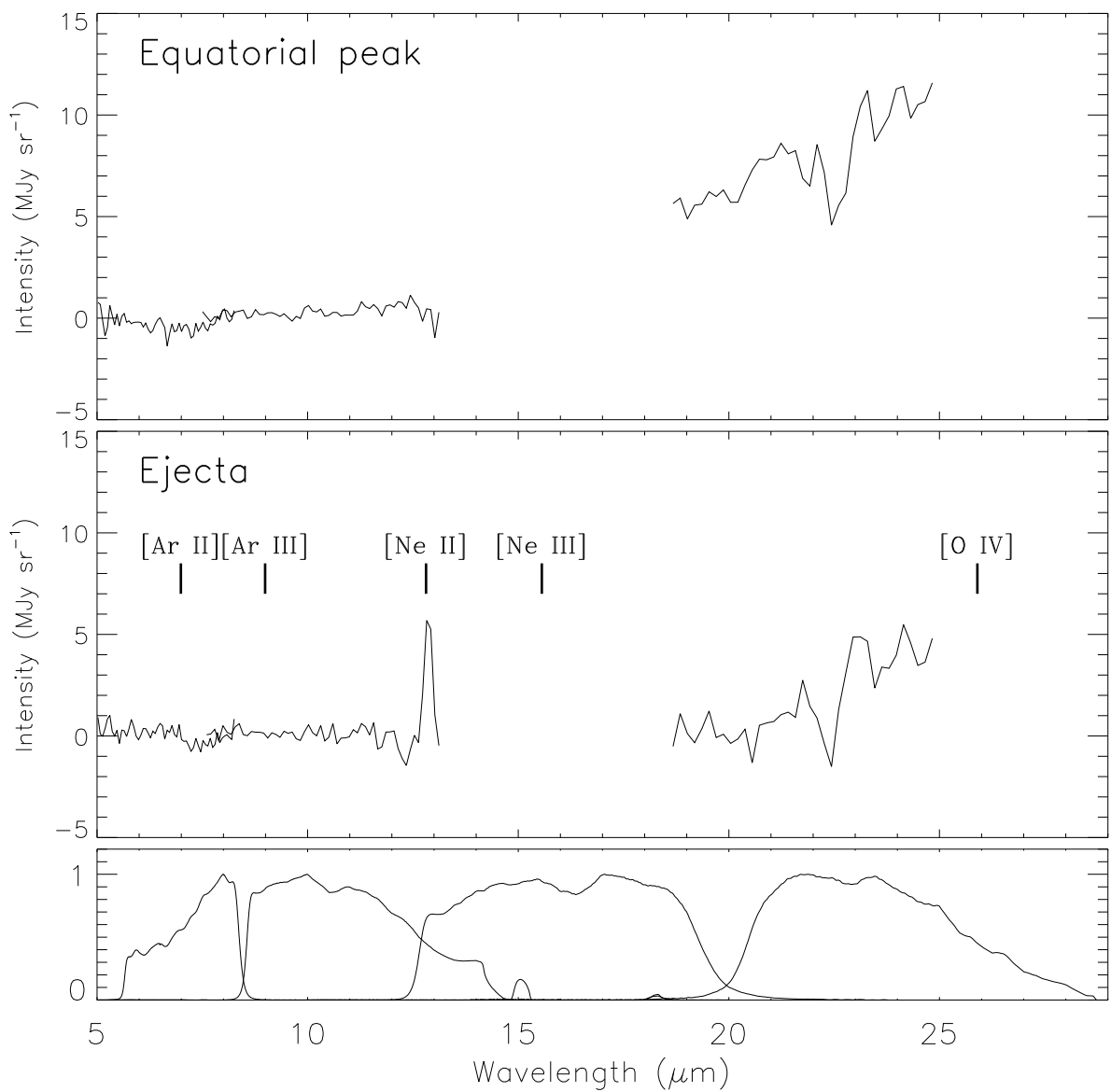}
\caption{The background subtracted spectra obtained at the 
equatorial peak and ejecta positions.
The MIR wavelength is covered by three elements:
SG1 (4.6--9.2 $\mu$m), SG2 (7.2--13.4 $\mu$m), and
LG2 (17.5--26.5 $\mu$m).
Note that the positions of LG2 spectra is slightly shifted from those of SG spectra.
Their positions and coordinates are indicated in Figure~\ref{fig_cmap}.
The relative spectral responses of S7, S11, L15, and L24 imaging bands
are indicated in bottom panel.
}
\label{fig_spec}
\end{figure}

Figure~\ref{fig_spec} shows the background-subtracted MIR grism spectra
of the peak positions of the ER and the IR-Ejecta.\footnote{We only 
present the MIR spectra because NIR spectra were heavily contaminated 
by emission from nearby field stars.}
(Note that the LG2 spectra at $>$ 17 $\mu$m were obtained from positions 
slightly shifted from those of the SG spectra of 5--14 $\mu$m as we described in \S~2.)
The spectrum from the IR-Ejecta peak clearly shows the [\ion{Ne}{2}] line at 12.8 $\mu$m,
which confirms the nature of the radiatively shocked ejecta.
No Ar lines (i.e., [\ion{Ar}{2}] at 7.0 $\mu$m and [\ion{Ar}{3}] at 9.0 $\mu$m) were detected,
while neither [\ion{Ne}{3}] line at 15.6 $\mu$m nor [\ion{O}{4}] line at 25.9 $\mu$m was covered.
Table~\ref{tab_line} lists the flux of the [\ion{Ne}{2}] line 
and upper limits of the [\ion{Ar}{2}] and [\ion{Ar}{3}] lines.
The flux of the [\ion{Ne}{2}] line at 12.8 $\mu$m  is 
2.8$\pm$2.5$\times$10$^{-5}$ erg cm$^{-2}$ s$^{-1}$ sr$^{-1}$,
corresponding to the surface brightness of 0.24$\pm$0.21 MJy~sr$^{-1}$ in the L15-band image.
Given the L15-band surface brightness of 1.6$\pm$0.1 MJy~sr$^{-1}$,
$\sim$~15~\% of the L15-band emission is due to the [\ion{Ne}{2}] line emission.
For the S11 band, the observed [\ion{Ne}{2}] line emission flux
corresponds to the S11-band surface brightness of 0.12$\pm$0.11 MJy~sr$^{-1}$.
This is equivalent to $\sim$~55~\% contribution to the total S11-band emission,
which is much larger than the case of the L15-band emission.
The difference is
because there is no strong emission line other than 
the [\ion{Ne}{2}] line in the S11 band
while there is additional [\ion{Ne}{3}] 15.5 $\mu$m line in the L15 band
(see the spectral responses in Figure~\ref{fig_spec}).
(There could be some dust emission too. See \S~4.3.)
On the other hand,
the spectrum at the ER peak is dominated by the continuum emission, although 
the LG2 spectra have a lower signal to noise ratio.

\subsection{Large-scale Infrared Emission around \snr}

\begin{figure}
\plotone{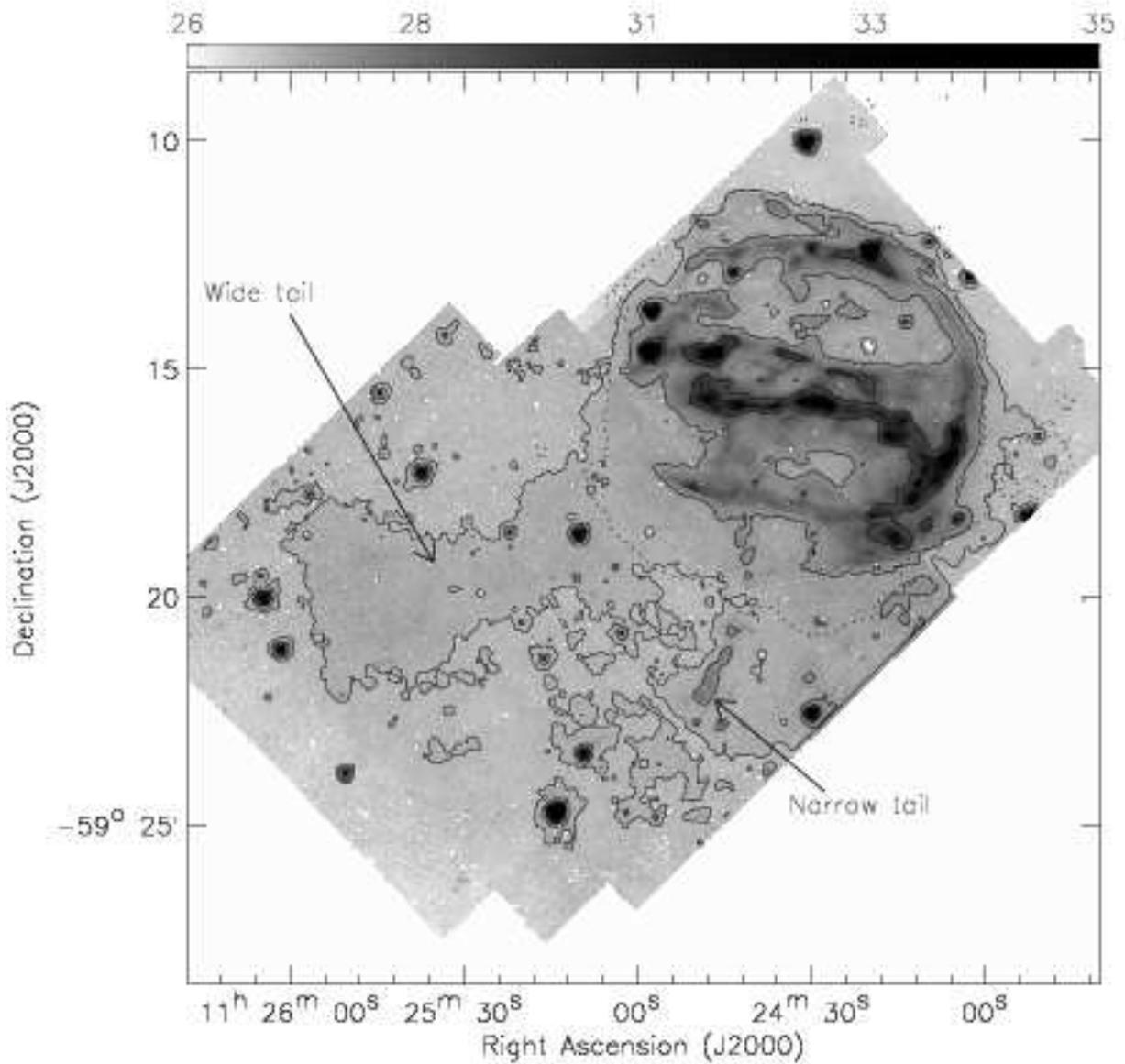}
\caption{The large-scale L18W-band image of G292.0$+$1.8.
Gray-scale range is indicated at the top.
Contour levels are
27.0, 27.5, 29, and 31 MJy sr$^{-1}$.
The dotted contour is the ATCA 20 cm radio-continuum boundary
\citep{gae03}.
The ``Narrow tail" and ``Wide tail" features outside shell boundary are labeled.
}
\label{fig_ircw}
\end{figure}

Figure~\ref{fig_ircw} presents a combined L18W-band image (\S~2) covering both the SNR and
an extended area in the southeast. 
Besides the features of the SNR that we already described in previous sections,
there is a notable feature of the L18-band emission in the south.
The elongated ``Narrow tail" is located around
($\rm 11^h24^m45^s$, $\rm -59^\circ22'00''$),
$\sim$~7\arcmin\ apart from the center of the SNR. 
It is close to the IR-Ejecta in the southeastern part of the SNR
and its elongation direction is roughly towards the center of the SNR.
Its extent is $\sim 1.5'$ or 2.6 pc.
In addition, there is faint, diffuse emission toward the south and southeast 
too. The emission toward the southeast appers distinct - it appears to protrude 
from the open portion of the OES having similar outer boundary with radio plateau
at the east and south, but extends $\sim 7'$ southeast beyond the radio boundary of the SNR.
We consider that this ``Wide tail" could be associated with the SNR. There is large, diffuse
\ion{H}{2} region superposed on G292.0$+$1.8  on the sky (RCW 54; Rodgers et al. 1960),
but its H$\alpha$ emission is elongated along the northeast-southwest direction without any emission
corresponding to the Wide tail in the Southern H-Alpha Sky Survey Atlas \citep{gau01}.
Therefore, the Wide tail is not associated with \ion{H}{2} regions and its association 
with the SNR is likely.

\begin{figure}
\plotone{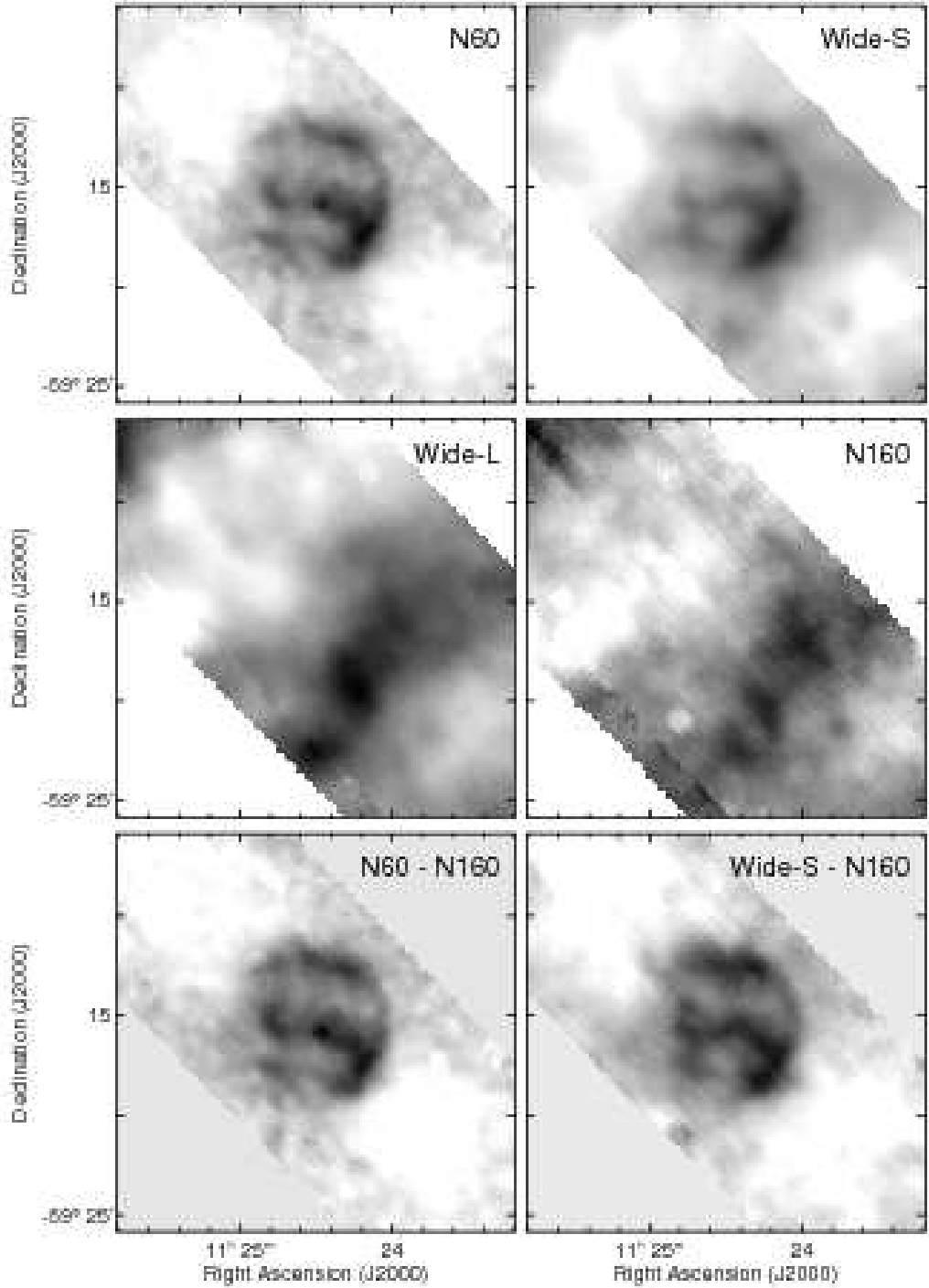}
\caption{$AKARI$ FIS images of G292.0$+$1.8. 
The scan direction is northeast to southwest.
The observed band is indicated at
the upper-right corner in each panel.
The background subtracted images are shown in the bottom panels.
The gray-scales of N60, Wide-S, Wide-L, N160, N60--N160, 
and Wide-S--N160 bands are
18--28, 32--44, 70--88, 52--64, 0--9, and 0--8 MJy sr$^{-1}$, respectively.
}
\label{fig_fis}
\end{figure}

Figure~\ref{fig_fis} presents the $AKARI$ FIS FIR-band images of G292.0$+$1.8.
The images cover the entire SNR with a scan direction of northeast to southwest.
The FIR emission of the SNR associated with the ER and OES
is clearly detected in the N60- and Wide-S-band images
with strong concentration in the southwestern part,
although fine details are difficult to see because of their low spatial resolutions.
The N160- and Wide-L-band images are, however, clearly different from images of the other bands.
They show an elongated feature extended in the northwest-southeast direction
which shows no correlation with the emission associated with the SNR.
Also its peak position is located outside the boundary of the SNR.
These indicate that the N160- and Wide-L-band emission
is not directly associated with
G292.0$+$1.8, although it is possible that the enhanced brightness in
the SW part of OES is due to the interaction of the remnant with this
extended structure \citep[cf.][]{bra86}.
The bottom panels of Figure~\ref{fig_fis} present 
the background-subtracted N60- and Wide-S-band images. 
The background emission was estimated by calculating scaling factors between the
N60- and N160-band images and also between the Wide-S- and N160-band images from the
areas outside the SNR.
Compared to the N60- and Wide-S-band images of the top panels,
the background-subtracted images show more clearly the emission
associated with the SNR, including the northern part of the OES
that is not clear in the orignial images.

\section{Discussions}

\subsection{Destruction of Circumstellar Dust}

\subsubsection{Infrared Emission from Shocked Circumstellar Dust}

The ER and OES are the most prominent features
in our $AKARI$ multi-band IR images (Figures~\ref{fig_irc} and \ref{fig_fis}).
The ER is composed of two filaments, where the southern one is brighter 
than the northern counterpart.
The southern filament appears to be of the normal composition in 
soft X-rays \citep{par02}
without any apparent radial motion in the optical \citep{gha05},
which led the authors to conclude that it is CSM from the progenitor of the SN 
in \snr.
We suggest that the northern filament, 
which has been clearly found by $AKARI$ observations, 
could be part of the same structure based on its similar distribution to 
the southern filament.  We note that the bright clump at      
($\rm 11^h24^m46^s$, $\rm -59^\circ14'42''$) near the eastern end of   
the northern filament has a counterpart in the [\ion{O}{3}] 
image of \citet{gha05}.
Its velocity is near zero, which supports that the northern and
southern filaments form a single structure. 
Furthermore, it is most clearly visible at the 
$-$120 km s$^{-1}$ frame of the Rutgers Fabry-Perot velocity scan images and
absent at the 0 km s$^{-1}$ frame where the southern filament is brightest 
(Figure 2 of Ghavamina et al. 2005). 
We interpret this velocity difference as an indication suggesting that
the northern and southern filaments are parts of a tilted, expanding
ring structure produced 
by the progenitor of the SN.
The location of the center point of the OES at the middle of the two filaments
also reconciles with the interpretation, given the CSM-nature of the OES (see below).
We note that such a, but smaller, ring structure of the CSM was found in SN 1987A and 
also possibly in the Crab nebula \citep{bou04, gre04}.
The OES, on the other hand, is relatively fainter in X-rays than the ER
\citep{par02}. 
The $AKARI$ MIR brightness and color of the OES, however,
are similar to those of the ER (Figure~\ref{fig_cmap}).
Toward the southern filament of the ER, 
\citet{gha05} reported the detection of the optical radiative lines
produced by the partially radiative shocks starting to develop cooling zones.
It implies that the X-ray emitting gas in the ER is cooler 
than that of the OES, while they have somewhat similar dust properties.
And there is a possibility that
the ER in S11-S7 difference image also contains the [\ion{Ne}{2}] line emission
produced in the CSM region where the shock has started to cool down to $<$ 100,000 K.
However, its contribution might be small, 
because we have not detected the [\ion{Ne}{2}] 12.8 $\mu$m line at the ER.

In addition to the ER and OES, there are faint MIR emissions 
beyond the northern and southeastern edges of the OES (Figure~\ref{fig_cmap}).
They have sharp outer boundaries, representing the current
locations of the SN blast wave. Faint X-ray emission was detected in
those areas too \citep{par02}. It is possible that the SN blast
wave has overtaken the OES and produced those emission features while
propagating into a more diffuse medium. On the other hand, it is also
possible that the remnant has a front-back asymmetry and they are just
projected boundaries of the more-extended shell. In any case, their
asymmetric spatial distribution 
suggests that either the ambient density distribution
or the mass ejection from progenitor was asymmetric.

\begin{figure}
\plotone{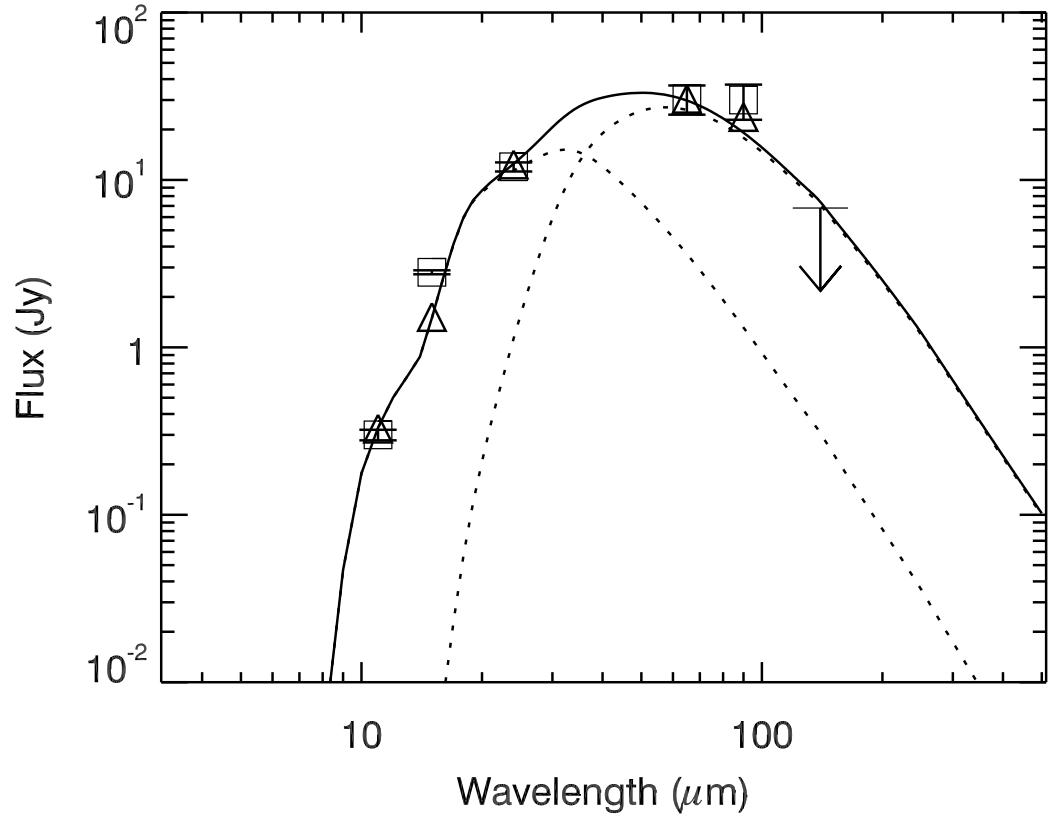}
\caption{The measured total fluxes and 
the best fit of SED using 
the modified blackbody composed of two temperature dust components.
The dust model is a mixture of carbonaceous and silicate
interstellar grains of $R_V$ = 3.1.
The boxes are the measured fluxes with errors
and the triangles indicate the color-corrected fluxes.
The downward arrow indicates the upper limit of 140 $\mu$m flux.
}
\label{fig_sedtot}
\end{figure}

As in Figure~\ref{fig_spec}, the MIR emission of \snr\ is dominated by continuum emission, 
not by line emission. 
This implies that the IR emission is 
from shock-heated dust grains in the CSM. 
Figure~\ref{fig_sedtot} presents the spectral energy distribution (SED) of the SNR 
in Table~\ref{tab_f}.
We applied modified blackbody fits composed of two dust components 
to the observed SED 
in order to obtain the best-fit dust temperatures.
The dust model based on a mixture of carbonaceous and silicate interstellar grain
of $R_{\rm V}$ = 3.1 \citep{dra03}
gave dust temperatures of $\sim$~103~K (warm dust) and $\simgt$~47~K (cold dust), 
corresponding to the mass of 4.5 $\pm$ 0.9 $\times$ 10$^{-4}$ M$_\odot$ 
and $\simlt$~4.8 $\times$ 10$^{-2}$ M$_\odot$, respectively.
(Note that the lower limit of the cold dust temperature comes from the 
upper limit of the flux at 140 $\mu$m.)
For the case of the dust model based on the graphite and silicate grain of 0.001 to 0.1 $\mu$m size 
\citep{dra84, lao93}, the derived total mass is 
in the range of (1.0--3.4) $\times$ 10$^{-2}$ M$_\odot$, 
comparable to the total mass obtained for the former model.
The derived dust mass corresponds to the dust-to-gas ratio of $\simlt$ 1.6 $\times$ 10$^{-3}$,
if we use the 30.5 M$_{\odot}$ swept-up mass (at distance of 6.2 kpc) 
of the gas in the CSM obtained 
in X-ray observations \citep{gon03}.
This is lower than the ratio 6.2 $\times$ 10$^{-3}$ found in the local ISM \citep{zub04}.
The low dust-to-gas ratios in the swept-up materials by the shock 
destruction were also obtained with Spitzer observations on the 
core-collapse SNRs in the Large Magellanic Cloud \citep{wil06}
and the Kepler \citep{bla07}. According to our result, 
$\simgt$ 75~\% of the dust in G292.0$+$1.8 might have been destroyed by 
the SN shock, or the initial dust-to-gas ratio surrounding G292.0$+$1.8
might be lower than the local value. It is comparable to the fraction 
derived in other SNRs, such as 64 \% in Cas~A,  \citep{dwe87} and 78 \% in 
Kepler \citep{bla07}.

\subsubsection{Model Calculations of Shock-heated Dust Emission}

We perform model simulations for the destruction of dust by SN shock waves
and the thermal emission from shock-processed dust.
The physical processes of dust in shocks have been so far discussed
in many works \citep[e.g.,][]{tie94, van94, dwe96, jon04}.
Once the circumstellar dust grains are swept up by the blast wave, they
acquire high velocities relative to the gas and are eroded by
kinetic and/or thermal sputtering in the shock-heated gas.
Dust grains are also heated by collisions with energetic electrons
in the hot gas and radiate thermal emission at IR wavelengths.
Dynamics, erosion, and temperature of dust depend on the temperature
and density of the gas as well as the chemical composition and size of
dust grains.

We adopt the model of dust destruction calculation by \citet{noz06},
in which the motion, destruction, and heating of dust in the
shocked gas are treated in a self-consistent manner by following the
time evolution of the temperature and density of the gas for the spherically 
symmetric shock wave.
As the initial condition of the SN ejecta we consider the freely expanding
ejecta with the velocity profile of $V=V_{\rm eje}(r/R_{\rm eje})$ and
the density profile of $\rho=\rho_{\rm core}$ at $r \le R_{\rm core}$ and
$\rho=\rho_{\rm core} (r/R_{\rm eje})^{-\alpha}$ at $r > R_{\rm core}$,
where $V_{\rm eje}$ and $R_{\rm eje}$ are the velocity and radius of the
outermost ejecta, respectively.
Taking the kinetic energy of $10^{51}$ ergs 
and the ejecta mass of 8 $M_{\odot}$ \citep{gae03}
for the density profile of core-collapse SNR of
$R_{\rm core} =0.3 R_{\rm eje}$ and $\alpha=12$ 
\citep[e.g.,][]{che82, mat99, pit01},
we obtain $\rho_{\rm core}=1.4 \times 10^{-19}$ g cm$^{-3}$,
$V_{\rm eje}=1.3 \times 10^9$ cm s$^{-1}$, and
$R_{\rm eje}=4.2 \times 10^{17}$ cm
at 10 yrs after explosion, when the simulations are started.
Since most of the dust
grains swept up by the forward shock during the later epoch of the evolution,
the calculation results are not sensitive to the ejecta structure.

For the ambient medium we consider the constant hydrogen number density of
 $n_{\rm H,0} =$ 0.1, 0.5, 1, and 10 cm$^{-3}$.
The circumstellar dust is assumed to be amorphous carbon or silicate
(forestrite) with the power-law size distributions
($\propto a^{-3.5}$) ranging from $a=0.001$ $\mu$m to $a=0.5$ $\mu$m
\citep{mat77}.
The optical constants are taken from \citet{edo83} and \citet{sem03}.
Based on the time evolution of the size distribution and temperature of
dust given by the simulation and the assumption of the dust grains being in
thermal equilibrium, we calculate the IR SED by thermal emission from
the shock-heated dust.
The detailed description for calculating the IR SED from shocked dust
will be given elsewhere (T.\ Nozawa et al.\ 2009 in preparation).

\begin{figure}
\plottwo{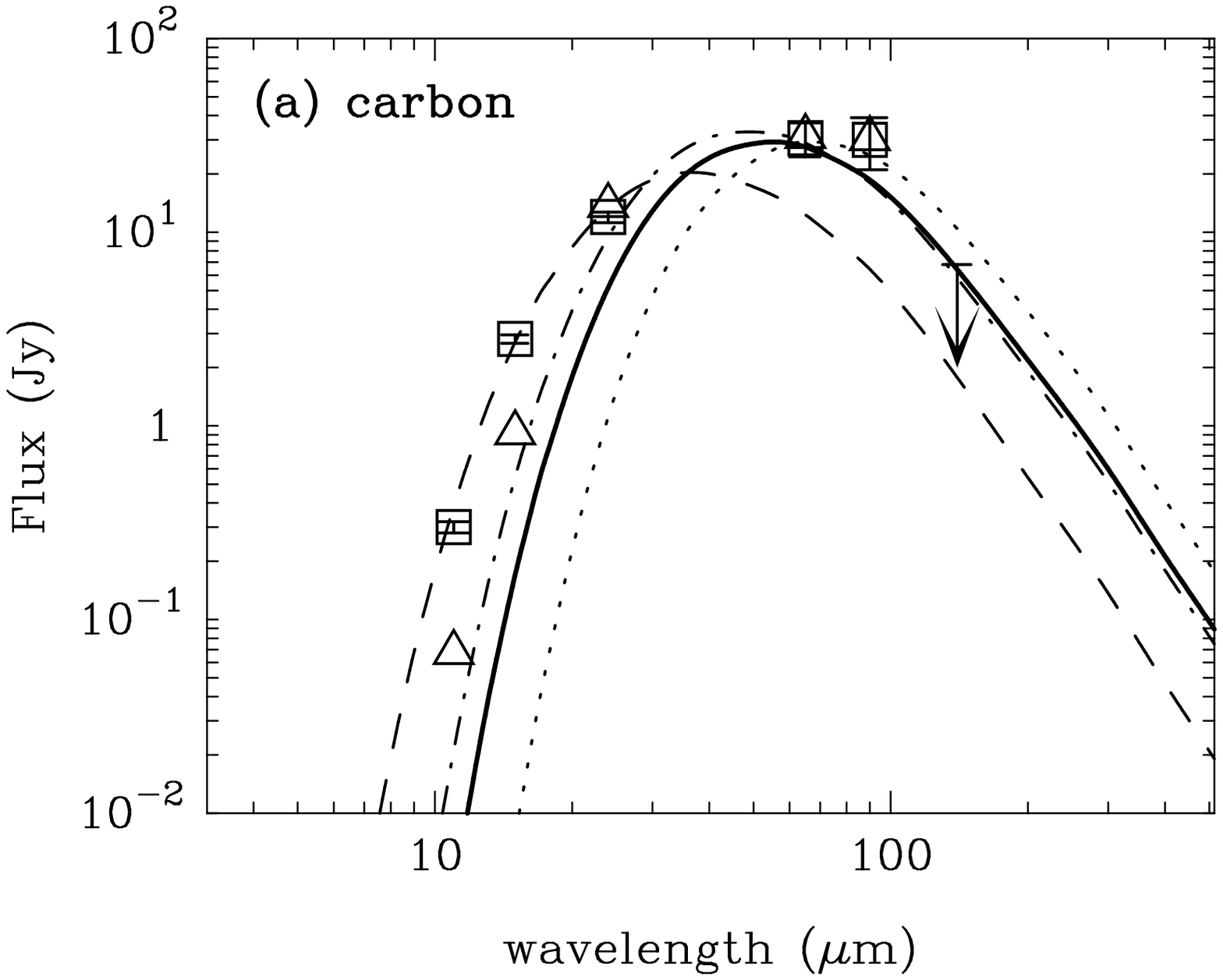}{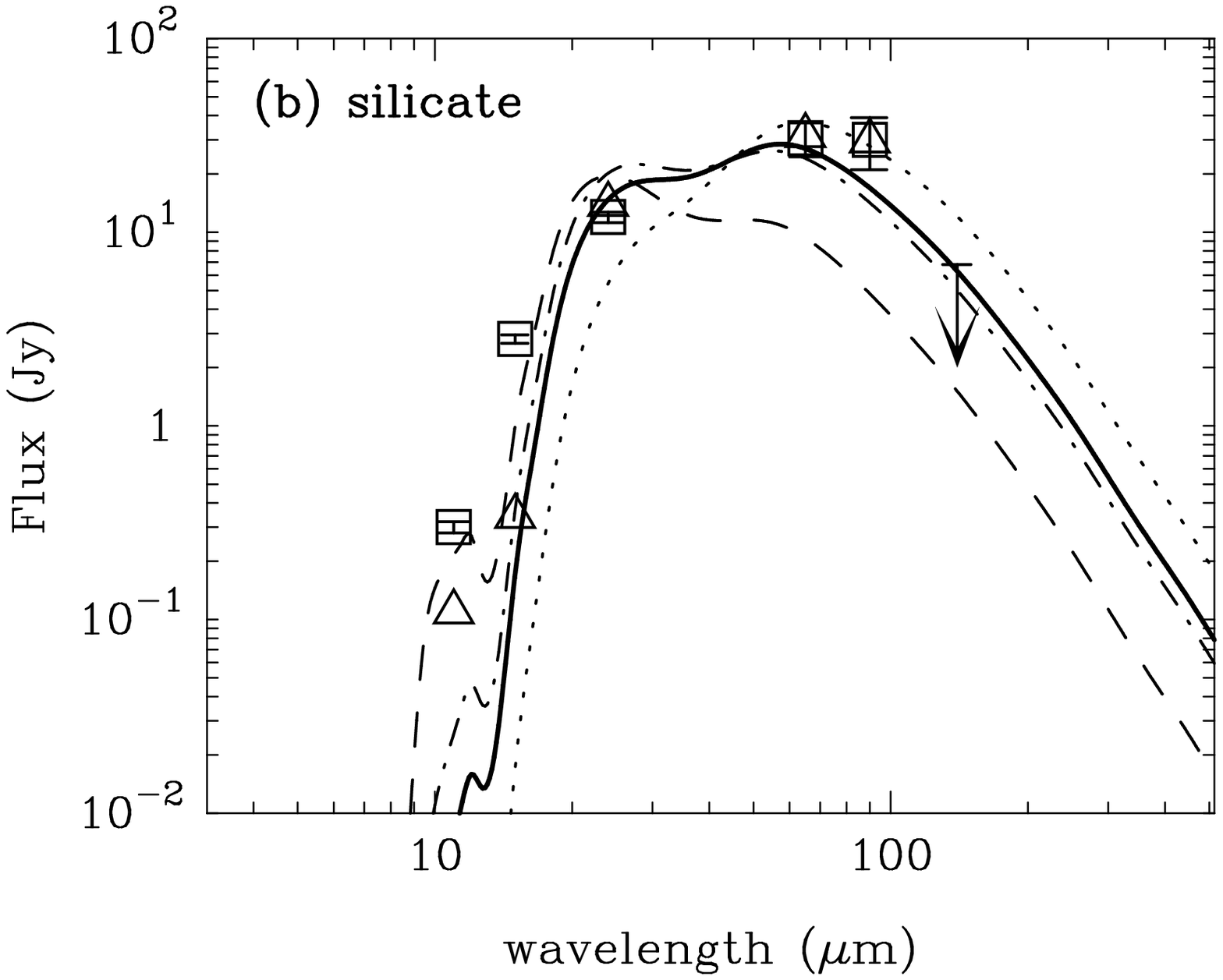}
\caption{
Comparison of the measured fluxes with the SED at 3,000 yrs 
obtained from the model calculations (See text for details.). 
The dust compositions are (a) amorphous carbon grains and
(b) silicate grains.
The dotted, solid, dashed-dotted and dashed lines indicate the results of simulations for
$n_{\rm H,0} =$ 0.1, 0.5, 1, and 10 cm$^{-3}$.
Symbols are the same as Figure~\ref{fig_sedtot}.
}
\label{fig_dmodel}
\end{figure}

Figure~\ref{fig_dmodel} compares the observed fluxes of G292.0$+$1.8 with
the calculated IR SEDs at 3,000 yrs for $n_{\rm H,0} =$ 0.1, 0.5, 1, and 10
cm$^{-3}$. We present the results with the initial dust-to-gas
mass ratio to best reproduce parts of the observed SED, for amorphous
carbon (Figure~\ref{fig_dmodel}a) and silicate (Figure~\ref{fig_dmodel}b).
The typical temperatures of dust are 35--55, 45--65, 50--70, and 60--80 K
for $n_{\rm H,0} =$ 0.1, 0.5, 1, and 10 cm$^{-3}$, respectively, and
the resulting dust masses are in the range
of (0.3--5) $\times 10^{-2}$ $M_\odot$ for carbon and
(0.4--8) $\times 10^{-2}$ $M_\odot$ for silicate with the higher values
for lower ambient density (thus lower temperature of dust).
It can be seen that
the results of the silicate grains with $n_{\rm H,0} =$ 0.5 cm$^{-3}$, 
which coincides with the density estimated from X-ray observations \citep{gon03},
can reasonably reproduce the overall shape
of the IR SED for G292.0$+$1.8.
In this case, the mass of grains
radiating IR emission is 
$2.9 \times 10^{-2}$ $M_\odot$.
Note that the initial dust-to-gas mass ratio corresponding this result is
$1 \times 10^{-3}$, which is smaller
than that in the local Galaxy.
If the IR emission originates in the swept-up materials 
containing the mass-loss wind of RSG
with the solar metallicity,
the condensation of silicate is expected,
and, according to our results, its condensation efficiency could be low.

It should be noted here that
the simulation results for $n_{\rm H,0} <$ 10 cm$^{-3}$ 
significantly underestimates the flux at short (11 $\micron$) wavelength.
However, the SED at shorter wavelengths could be
resolved by including the effect of a stochastic heating of small grains;
stochastically heated dust produces more emission at shorter
wavelengths than the dust with equilibrium temperature, and it may also
allow acceptable fits for lower initial density and/or
different size distribution of dust than the current
best fit.
Alternatively, this disagreement may be caused by the difference in the
assumed composition of dust.
To gain deeper insights into the properties of dust in
G292.0$+$1.8, we need further investigations by taking account of the
stochastic heating and changing the composition and size distribution of
dust as well as the density profile in the ambient medium.

\subsection{Infrared Emission from Supernova Ejecta}

\subsubsection{Ejecta Neon Line Emission}

We have detected MIR emission associated with the SN ejecta: IR-Ejecta. 
It is prominent in the 15/24 $\mu$m ratio image by its high L15/L24 ratio 
(Figure~\ref{fig_cmap}), 
but is marginally seen in the total intensity maps 
of the MIR (11--24 $\mu$m) and FIR (65--90  $\mu$m) too. 
The emitting area coincides with the fast-moving ($< 1,500$ km s$^{-1}$), 
O-rich SN ejecta: the triangle-shaped bright portion near the equatorial filament 
was called ``spur" and the extension to the south was called ``streamers" 
by \citet{gha05}. 
In the high-resolution [\ion{O}{3}] 5007 image, 
the spur is crescent-shaped with a sharp boundary toward the SNR center, 
while the streamer appears to be composed of clumps embedded 
in diffuse emission. 
Recent measurement of their proper motions showed that 
they are expanding systematically from a point near the 
geometrical center of the OES \citep[][see \S~4.4 too]{win08}.

According to our spectroscopic result, 
the IR-Ejecta shows \neontwo\ 12.8 $\mu$m emission 
but no [\ion{Ar}{2}] 7.0 or [\ion{Ar}{3}] 9.0 $\mu$m emission. 
Note that the latter lines from Ar ions were also detected 
at the metal-rich ejecta in young SNRs 
\citep{dou01, smi08, wil08}. 
The lack of IR line emission from Ar ions in this area 
is consistent with the results of optical or X-ray studies 
which showed that there is no line emission from elements 
heavier than S in this area 
(Ghavamian et al. 2005 and references therein; Park et al. 2002). 
The absence of Ar lines supports the claim by 
\citet{gha05} that we are not seeing the ejecta in the inner-most region 
accelerated by pulsar wind nebula 
but seeing the He-burning-synthesized, 
O-rich ejecta swept-up by reverse shock.   

The IR-Ejecta shows high L15/L24 ratio, i.e., L15/L24=0.88
compared to 0.25 of the shocked CSM.
The high L15/L24 ratio is at least partly due to Ne lines. 
(There is [\ion{O}{4}] 25.9 $\mu$m line in the L24 band, but it is weak, i.e.,
$\sim 20$\% of the [\ion{Ne}{2}]+[\ion{Ne}{3}] lines. See \S~4.3.)
According to our estimation, \neontwo\ 12.8 \mum\
line contributes $\sim$~15~\% to the L15 flux. 
In the L15 band, there is another strong Ne line,
\neonthree\ 15.6 \mum\ line. 
We may estimate the possible contribution of \neonthree\ 15.6 \mum\ line
in G292.0+1.8 as follows.

The strength of a forbidden Ne line is given by
$I_{21}=(h\nu_{21}/4\pi)N_2 A_{21}$ 
where $\nu_{21}$ is the frequency of the line,
$N_2$ is the column density of Ne$^+$ or Ne$^{++}$ ions 
in the upper state along the line of sight, 
and $A_{21}$ is the Einstein coefficient, 
so that the \neonthree\ 15.6 \mum/\neontwo\ 12.8 \mum\ ratio is given by
\begin{equation}
{I_{15.6~\mu m} \over I_{12.8~\mu m}}
= 0.57 {N({\rm Ne}^{++}) \over N({\rm Ne}^+)}{f({\rm Ne}^{++}, {^3}P_1) 
\over f({\rm Ne}^{+}, {^2}P_{1/2})} ~,
\end{equation}
where $N({\rm Ne}^{++})$ and $N({\rm Ne}^{+})$ are column densities 
of Ne$^{++}$ and Ne$^{+}$ ions,
$f({\rm Ne}^{++}, {^3}P_1)$ and $f({\rm Ne}^{+}, {^2}P_{1/2})$ 
are the fractions of Ne$^{++}$ and Ne$^{+}$ ions 
in the upper states of the corresponding emission lines
(see Glassgold et al. 2007 for atomic parameters of these lines).
Assuming statistical equilibrium, 
the last factor, ${f({\rm Ne}^{++}, {^3}P_1) /f({\rm Ne}^{+}, {^2}P_{1/2})}$,
can be easily calculated \citep[e.g.,][]{gla07}, and
varies from 1 to 2 as the electron density increases from a low-density limit
to densities much higher than the critical density
($=(4-6)\times10^5$~cm$^{-3}$ at 10,000 K).
Therefore, the intensity ratio $I_{15.6~\mu m}/I_{12.8~\mu m}$
is $\simlt 1$ unless $N({\rm Ne}^{++}) \gg N({\rm Ne}^+)$, which
happens at temperatures $\simgt 10^5$~K in collisional equilibrium
\citep[e.g.,][]{all69}.
\citet{gha05} suggested that ejecta undergoes radiative shocks 
with a velocity of  50--200 km s$^{-1}$
from the [\ion{O}{3}] 5007 line widths.
The temperature of shocked ejecta
gas immediately behind the shock of velocity $v_s$
should be high, i.e.,
$4.5\times 10^6~(v_s/100~{\rm km s^{-1}})^2$~K,
assuming pure neon preshock gas, but since the shocked gas element
cools fast due to enhanced heavy elements and the column density
of the gas at $T$ might vary roughly with $1/T$, 
we expect $N({\rm Ne}^{++})< N({\rm Ne}^+)$ and therefore
${I_{15.6~\mu m}/I_{12.8~\mu m}}\simlt 1$. 
Toward the SN ejecta in young core-collapse SNRs such as Crab,
Cas A, the SNR B0540$-$69.3 in the LMC, or the SNR 1E0102$-$72.3 in the SMC,
the \neonthree 15.6 \mum/\neontwo 12.8 \mum\ ratio
varies from 0.3 to $\sim 1$ too \citep{dou01, tem06, smi08, wil08, rho09}. 
We may conclude that the contribution of 
the \neonthree 15.6 \mum\ line to the L15 band should be
at most comparable to that of the \neontwo 12.8 \mum\ line 
unless the shock is truncated at
relatively high temperatures.

\subsubsection{Dust Emission Associated with Ejecta}

We have derived IR fluxes associated with the ejecta 
and the results are summarized in Table~\ref{tab_f}.
It is difficult to extract the ejecta emission in IR observations,
although our results unveiled the ejecta region unambiguously.
The IR-Ejecta is clearly divided into two regions 
(Figures~\ref{fig_cmap} and \ref{fig_cplot});
one that
coincides with the ejecta-only region and has a tight correlation in
the L15 vs L24 plot and the other
superposed on the ER and has a loose correlation in the L15 vs L24 plot.
These IR-Ejecta regions are part of the structure 
defined as the [\ion{O}{3}] ``Spur" by \citet{gha05}, 
with the ejecta-only region being the bottom of the Spur, 
and the ejecta superimposed
on the ER being the top portion of the Spur.
We measure the flux at the ejecta-only region 
masking the emission from stars.
The area of the ejecta-only region is more than four times larger than
the superposed ejecta region 
and it contains most of the flux.
In order to remove the background emission,
we subtract the average of nearby ($\sim$ 2$'$) low intensity region.

The $AKARI$ ejecta spectrum in Figure~\ref{fig_spec}
shows that the 11 and 15 $\mu$m fluxes are not from dust continuum. 
We believe that 
the S11-band is mostly dominated by line emission,
while the L15-band needs an additional component (see \S~3.3).
In the next paragraph, 
we show that some of the L15 flux is due to a bump at
15--25 $\mu$m. 
On the other hand, FIR emission is certainly dominated by dust 
continuum. The L24-band contains the [\ion{O}{4}] 25.9 $\mu$m line emission, 
whose wavelength is not covered with our spectroscopic 
observations. Its contribution in L24-band is expected
to be less than 20 \% in our estimation using the [\ion{O}{4}] 25.9 $\mu$m 
line intensity in \citet{gha09}. 
At first, we fit 24--140 $\mu$m band fluxes using
a modified-blackbody for silicate and graphite grain models of 0.001--0.1
$\mu$m size \citep{dra84, lao93}. The best fits give
temperature of 64--88 K and dust mass of 2.0--8.2 $\times$ 10$^{-5}$
M$_\odot$. 
If we exclude the L24-band data point, 
the temperature drops to 44--52 K and 
the dust mass increases by a factor of 5. 
This IR-emitting dust mass is much smaller ($<$ 4 $\times$ 10$^{-4}$)
than the theoretically predicted dust mass (0.1--1 M$_\sun$) 
to be formed in the core-collapse SN explosion \citep[e.g.,][]{noz03}.

\subsection{Comparison with Spitzer Spectroscopic Results}

Recently the results from MIR observations of \snr\ 
with the $Spitzer$ satellite have been published \citep{gha09} 
based on low-resolution spectroscopy of the ejecta and the southern filament of the ER
where the ejecta emission is superimposed on the emission from the CSM.
The former is located $\sim$~0\arcmin.5 north from our IR-Ejecta slit position,
while the latter, which has a high L15/L24 ratio,
is $\sim$~1\arcmin\ apart in the northeast direction.
Their ejecta spectrum shows strong emission from the
[\ion{Ne}{2}] 12.8 $\mu$m, [\ion{Ne}{3}] 15.6 $\mu$m,
[\ion{O}{4}] 25.9 $\mu$m lines and relatively weak 
[\ion{Ne}{5}] 24.4 $\mu$m and [\ion{Ne}{3}] 36.0 $\mu$m lines,
but no lines of Mg, Si, S, Ar or Fe are identified.
(Note that the Si and S lines in their spectrum are not from 
the ejecta but from the background.)
The observed ratio of [\ion{Ne}{2}] 12.8 $\mu$m to [\ion{Ne}{3}] 15.6 $\mu$m is 2.1--2.7. 
The non-detection of Ar lines and the observed ratio of Ne lines are 
consistent with our results and the prediction in \S~4.2. 
The observed flux of [\ion{O}{4}] 25.9 $\mu$m line is $\sim 20$\% 
of the [\ion{Ne}{2}] plus [\ion{Ne}{3}]
line fluxes \citep{gha09}. If the $AKARI$ L15 and L24 fluxes are entirely due to lines, 
we estimate that the L15/L24 ratio should be 2.3--2.6 using the line fluxes in
\citet{gha09}. (It becomes higher if we consider the 15-25 $\mu$m bump 
in the next paragraph.) The observd ratio, however, is 0.25 for the ER and OES 
and 0.88 for the ejecta. Therefore, the contribution of
[\ion{O}{4}] 25.9 $\mu$m line flux to the observed L24-band flux should be small,
particularly toward the ER where $I_\nu(24)/I_\nu(15)\approx 5$ (\S~3.2).
We, however, note that the $Spitzer$ IRS spectrum toward the ejecta in 
\citet{gha09} does not show any obvious dust continuum emission 
between 24 and 36 $\mu$m. We consider that it could be because there was 
dust continuum emission in the IRS background. In the $AKARI$ 24 $\mu$m image, 
there is faint filament coincident with the IRS background region 
(LL1 Sky in Ghavamian et al. 2009). This filament is associated with 
the OES and its 24 $\mu$m emission might be dominated by dust continuum, 
so that, by subtracting its spectrum from the ejecta spectrum, 
the continuum feature could have been removed.

An interesting feature in the $Spitzer$ ejecta spectrum is a weak 15--25 $\mu$m bump,
which was suggested to be produced by newly-formed dust or swept-up PAHs along the line of sight. 
First of all, this bump explains the high L15/L24 ratio at the ejecta region described in \S~4.2.1.
The peak of the bump feature appears in the L15 band.
Therefore, it contributes mainly to the $AKARI$ 15 $\mu$m band.
Secondly, our $AKARI$ observations show that the areas with a high L15/L24 ratio
are coincident with those of the optical O-rich ejecta region. 
This supports the interpretation that the bump feature is related to the newly-formed dust 
in association with the SN ejecta, not to the swept-up PAHs.

According to \citet{gha09},
the $Spitzer$ spectrum from the southern filament of the ER is consistent with 
the emission from two dust components:
a warm (or hot) component of $\approx$ 114 K 
and a cold component of $\approx$ 35 K.
The temperature of the warm dust is higher than what we estimated with the $AKARI$ data
by $\sim$~10~K, while that of the cold dust is lower than 
the $AKARI$ temperature by a comparable amount.
It is possible that the presence of the bump feature in the $Spitzer$ spectrum resulted in
a higher temperature for the warm dust.
For the cold dust temperature, as explained in \citet{gha09},
the absence of the longer wavelength ($>$ 30 $\mu$m) data in the $Spitzer$ spectrum
is likely the primary reason of the lower temperature compared to the $AKARI$ results.
IR spectroscopic observations covering a broad band are necessary to clearly resolve 
the issues.

\subsection{Supernova Explosion in \snr}

\subsubsection{Circumstellar Shell and the Explosion Location}

One interesting result that we obtained in this study is 
the difference ($\sim$ 42\arcsec\ $\simeq$ 1 pc)
between the center of the CSM and the dynamical center of the ejecta (Figure~\ref{fig_cmap}).
The former corresponds to the center of the ER and OES that we determined with the $AKARI$ results;
the latter was determined by the distribution of the O-rich ejecta in the optical,
which is close to the center of the SNR in the radio emission.
In addition, the position of the pulsar is different from the both positions --
it is shifted in the southeast direction from the dynamical center by $\sim$ 46\arcsec.

One possibility is that the progenitor star was at the center of OES
during its RSG phase but exploded at the position of the dynamical
center of optical knots. This is possible because the RSG wind could
be confined by external pressure while the central star is moving.
According to \citet{che05}, the RSG wind from a 25--35 M$_\odot$ star,
which explodes as SN IIL/b, would be pressure confined while its outer
radius becomes $\simgt 5$ pc. The radius of OES (6 pc) is comparable
to what the theory predicts. If the progenitor star was moving at 
$\sim$ 10 km s$^{-1}$
and exploded after $\sim$ 10$^5$ yrs of the pressure confinement of the
shell, then the explosion center would be close to the dynamical
center of optical knots.

Another possibility is that the SN
exploded at the center of ER and OES, not at the dynamical
center of the ejecta or the center of the radio emission. 
This is motivated by the fact that the OES shows a very well-defined shell
structure surrounding the SNR, thereby the center of the OES may pinpoint to the real
location of the progenitor, which later exploded as a SN. 
On the other hand, the centers of
the radio emission and the O-rich ejecta motion could have been
weighted toward the southeast: 
Firstly, the geometrical center of the radio nebula could be significantly 
weighted toward the southeast because of the existence of the bright PWN.
Secondly, the dynamical center of optical knots is also weighted 
to the southeast because the bright optical knots in the southeastern
area are moving relatively slowly 
and the center is derived using an assumption of 
an unhindered constant velocity since the explosion
\citep{win08}.

If the SN explosion in \snr\ indeed occurred at the center of the OES, 
then the tangential velocity of the pulsar needs to be $\sim$ 1,000 km s$^{-1}$.
Although 1,000 \kms\ is somewhat large as a pulsar kick velocity, 
it is still within the acceptable range \citep[e.g.,][]{ng07}, 
and the pulsar in another O-rich SNR Puppis~A may also have such a high velocity \citep{hb06}.

\subsubsection{Ejecta Distribution and Explosion Asymmetry}

The optical and X-ray studies of the SN ejecta in G292.0$+$1.8 have shown that 
their spatial distribution and physical properties are not symmetric,
including the northwest-southeast concentration of the O and Ne ejecta,
and the distinctive difference of the X-ray plasma temperature between the 
northwestern and southeastern areas \citep{par02, gha05, par07}. 
Also the ejecta in the northern and southern boundaries move faster than
those in the eastern and western areas \citep{win08}. 
It is worth to note that the pulsar jet axis is also along the
northeast-southwest direction \citep{par07}.

Our $AKARI$ result on the Ne-line emitting ejecta material 
identified by their high L15/L24 ratio 
is consistent with the spatial distribution seen in [\ion{O}{3}] 5007: 
most of the ${\rm Ne}^+$/${\rm Ne}^{++}$ ions are distributed in the southeastern area 
called spur and streamers, 
several isolated ones coincide with the optical knots,
and the other group of Ne-rich knots is distributed in the northwestern area
and coincides with [\ion{O}{3}] optical emission knots
(fast-moving knots, or FMKs, as reported by \citet{gha05} and \citet{win06}).
The northwest ejecta is less obvious in optical
but it is easily recognized in the X-ray image of 
ionized O and Ne elements \citep{par02}.
This type of bipolar ejecta distribution 
is also found in other young core-collapse SNRs, e.g., 
Cas A and G11.2$-$0.3 \citep{smi08, koo07, moo09}.
Cas A shows a similar Ne ejecta distribution, which is almost 
perpendicular to the well-known narrow northeast-southwest jet,
but is roughly aligned to the bipolar ionic ejecta, 
which sometimes is suggested be the major direction of explosion
\citep{hwa04, whe08, smi08}. 
In G11.2$-$0.3, which is the remnant of the historical SN AD 386, 
the iron ejecta is found to be distributed mainly along northwest-southeast direction
\citep{koo07, moo09}.

The symmetry axis in the spatial and kinematical 
distribution of ejecta in G292.0$+$1.8, therefore, is either 
along northwest-southeast or north-south, which is perpendicular to the plane 
of the ER. 
We consider that either the explosion was asymmetric and/or 
the CS wind was denser in the equatorial plane so that the ejecta  
expanding in this plane was slowed down more 
compared to those expanding to the other directions. 

An interesting feature is the Narrow tail in the wide L18W-band image 
that extends from the end of the streamers to outside the remnant. 
A possible explanation is that the Narrow tail is a part of the O-rich streamer.
Note that the Narrow tail is connected to the streamers 
by a southern patch of emission near the boundary of L15-band image.
The distance from the center to the end of the Narrow tail is 7$'$ (13 pc),
which is $\sim$ 2$'$ (3 pc) farther than the outermost O-rich clump in this area
\citep{win08}. 
It, however, does not has an optical counterpart in the [\ion{O}{3}] image \citep{win08}.
If we assume a constant velocity and adopt an age of 3,000 yrs 
\citep{gha05, win08},
the transverse velocity of the Narrow tail is 4,000 km s$^{-1}$.
This gives a possibility that its radial velocity is also very large,
beyond the velocity range of previous optical narrow-band imaging observations
\citep[e.g., $\sim$ 2,000 km s$^{-1}$ in][]{win06}. 
An alternative explanation is that 
the Narrow tail is a reradiated IR light echo similar to 
the echoes identified in Cas~A \citep{kra05, dwe08}.
The lack of counterpart 
in the optical [\ion{O}{3}] and X-ray images,
together with the faint feature in the FIR images (Figure~\ref{fig_fis}), 
could be consistent with the continuum origin of the IR emission.
In case of a light echo, the location of the echo can be derived from 
the geometrical equation of ellipse 
whose two focuses are the SN and the observer \citep[e.g.,][]{cou39, dwe08}. 
Applying a projected radius of 13 pc
and a time delay of 3,000 yrs,
we obtain a location of the IR echo at 450 pc behind G292.0$+$1.8
and its angular offset of $\sim$ 2$^\circ$ from the line of sight.
It appears that this alignment is too tight at a first glance,
but it can be a selection effect caused by our limited imaging area.
For example, \citet{kra05} discovered IR echoes along the scan direction in Cas~A
for the first time, but \citet{dwe08} reported that 
the echoes were distributed in many positions 
in large ($\sim$ 2$^\circ$) area.
More observations
are definitely necessary to inspect the nature of the Narrow tail.

On the other hand, 
the sharp boundary of the SN blast wave 
was detected only toward the north and southwest (Figure~\ref{fig_cmap}). 
The absence of SN blast wave in the other directions implies 
that either the shock is trapped within the shell 
because that part of the shell has a larger column density 
or the SN blast wave has propagated far beyond 
because the ambient density is lower toward that direction. 
There is no indication in the $AKARI$ images 
that the OES is denser, where the SN blast wave is missing. 
Instead those parts are fainter in the FIR images 
which indicates a lower column density. 
In this regard, the faint MIR emission 
that extends far beyond the bright shell to the southeast (Figure~\ref{fig_ircw}) 
is interesting because if it is part of the SNR, 
it indicates that the SN blast wave has propagated much further out 
toward this direction probably due to the lower ambient density. 
A deep radio observation could reveal faint features 
associated with this IR structure.

\section{Conclusions}

We have presented the NIR to FIR imaging 
and MIR spectroscopic observations 
of the O-rich SNR G292.0$+$1.8 
using the IRC and FIS instruments aboard $AKARI$ satellite.
The almost continuous multiband imaging capability of $AKARI$ 
covering wide IR wavelengths together its wide field of view
enabled us to clearly see the distinct IR emission 
from the entire SNR and to derive its IR charactersitics.
We derive the physical parameters of IR-emitting dust grains 
in the swept-up circumstellar medium
and compare the result with the model calculations 
of dust destruction by a SN shock.
The overall shape of the observed SED can be explained
by a simple model using characteristic SNR parameters 
with a bit lower initial dust-to-gas ratio. 
At 11 $\mu$m, the model flux is significantly smaller, 
which may indicate the importance of stochastic heating.   
We have not detected any signficant amount of freshly-formed dust 
associated with the SN ejecta. 

The AKARI results in this paper give new insights
into the explosion dynamics of G292.0$+$1.8. 
We have discovered an almost symmetric IR shell 
probably produced by the circumstellar wind 
from the progenitor star in the RSG phase. 
Its center is significantly offset from 
the previously suggested explosion centers. 
We consider that either OES
represents the circumstellar shell pressure-confined by external medium 
or the SN exploded close to the center of the  OES. In the
latter case, 
the pulsar in G292.0$+$1.8 may 
be traveling at a speed of $\sim 1,000$~\kms.
At the same time,
the ejecta distribution is unveiled by their 
high 15 to 24 $\mu$m ratio.
The ejecta are mainly distributed along the northwest-southeast direction.
This symmetric pre-supernova structure and 
asymmetric ejecta distribution appear to be rather common 
in the remnants of SN IIL/b, which suffer strong mass-loss like Cas A
\citep{hin04, smi08}.
There is also a Narrow tail outside the SNR shell, 
which might be similar to the feature observed in Cas~A.
A detailed study is necessary 
to understand the nature of this feature.

We suggest that 
multi-band IR imaging observations are 
powerful tools to explore both the 
ejecta and CSM emission in young core-collapse SNRs.
Especially the 15 and 24 $\mu$m images are useful to reveal
the detailed structure of IR features,
which leads to better understanding of environments of the progenitor 
and the SN explosion.

This work is based on observations with $AKARI$, a JAXA project with the participation of ESA. 
We wish to thank all the members of the $AKARI$ project. 
We also thank B. Gaensler for providing the ATCA 20 cm image
and S. Park for providing the $Chandra$ X-ray image.
This work was supported by the Korea Research Foundation Grant funded by the Korean Government
(KRF-2008-357-C00052) and
the Korea Science and Engineering Foundation (R01-2007-000-20336-0).
This work was also supported in part by a Grant-in-Aid for Scientific
Research for the Japan Society of Promotion of Science (18204014).
T.N.\ has been supported in part by World Premier International 
Research Center Initiative (WPI Initiative), MEXT, Japan, and by the
Grant-in-Aid for Scientific Research of the Japan Society for the
Promotion of Science (19740094).

{\it Facility:} \facility{Akari}

\clearpage

\clearpage

\clearpage

\clearpage

\clearpage

\clearpage

\clearpage

\clearpage

\clearpage
\begin{deluxetable}{l ccl}
\tabletypesize{\scriptsize}
\tablewidth{0pt}
\tablecolumns{4}
\tablecaption{Summary of $AKARI$ observations
\label{tab_obssum}}
\tablehead {
\colhead{Date} &\colhead{ID\tablenotemark{a}} &\colhead{Mode} &\colhead{Band}
}
\startdata
2007. 1. 17.      & 1400748     & IRC imaging     & N3, N4, S7, S11 \\
2007. 1. 17.      & 1400749     & IRC imaging     & L15, L24   \\
2007. 1. 19.      & 1400751     & FIS imaging     & N60, Wide-S, Wide-L, N160 \\
2007. 7. 21.      & 1402801     & IRC imaging     & L18W \\
2007. 7. 20.      & 1402839, 41, 43 & IRC spectroscopy & NIR, MIR-S \\
2007. 7. 21.-22. & 1402840, 42, 44 & IRC spectroscopy & MIR-L
\enddata
\tablenotetext{a}{$AKARI$ observational identification number.
These observations are part of $AKARI$ mission program,
interstellar medium in our Galaxy and nearby galaxies (P.I.: H. Kaneda).
}
\end{deluxetable}

\clearpage
\begin{deluxetable}{lc cccccc cccc}
\tabletypesize{\scriptsize}
\tablewidth{0pt}
\tablecolumns{11}
\tablecaption{Characteristics of $AKARI$ imaging observations 
\label{tab_obsimg}}
\tablehead {
\colhead{Instrument} &\multicolumn{6}{c}{IRC}  &\multicolumn{4}{c}{FIS} 
\\
\cline{1-2} \cline{2-7} \cline{8-11}
\colhead{Channel} &\multicolumn{2}{c}{NIR} &\multicolumn{2}{c}{MIR-S}
&\multicolumn{2}{c}{MIR-L}
&\multicolumn{2}{c}{SW} &\multicolumn{2}{c}{LW}
\\
\colhead{} &\multicolumn{2}{c}{512$\times$412 InSb} 
&\multicolumn{2}{c}{256$\times$256 Si:As}
&\multicolumn{2}{c}{256$\times$256 Si:As}
&\multicolumn{2}{c}{Ge:Ga} &\multicolumn{2}{c}{stressed Ge:Ga} 
}
\startdata
Band &N3 &N4 &S7 &S11 &L15 &L24
&N60 &Wide-S &Wide-L &N160 \\
Reference wavelength ($\mu$m)
& 3.2 & 4.1 & 7.0 & 11.0& 15.0& 24.0
& 65  & 90  & 140 & 160 \\
Effective bandwidth  ($\mu$m)
& 0.87& 1.53& 1.75& 4.12& 5.98& 5.34
& 21.7& 37.9& 52.4& 34.1\\
Pixel size           ($''$)  
& 1.46& 1.46& 2.34& 2.34& 2.51$\tablenotemark{a}$& 2.51$\tablenotemark{a}$
& 26.8& 26.8& 44.2& 44.2\\
FWHM                 ($''$)  
& 4.0 & 4.2 & 5.1 & 4.8 & 5.7 & 6.8 
& 37  & 39  & 58  & 61 
\enddata
\tablenotetext{a}{Represents the width of pixel. 
The exact size is 2$''$.51$\times$2$''$.34.}
\end{deluxetable}

\clearpage
\begin{deluxetable}{l cccc}
\tabletypesize{\scriptsize}
\tablewidth{0pt}
\tablecolumns{5}
\tablecaption{Characteristics of $AKARI$ spectroscopic observations
\label{tab_obssp}}
\tablehead {
\colhead{Channel} &\colhead{NIR} &\multicolumn{2}{c}{MIR-S} &\colhead{MIR-L}
}
\startdata
Grism & NG &SG1 &SG2 &LG2
\\
Slit size ($''\times'$)
& 5\arcsec~$\times$~0.\arcmin8 & 5\arcsec~$\times$~0.\arcmin8 
& 5\arcsec~$\times$~0.\arcmin8 & 7\arcsec~$\times$~0.\arcmin4
\\
Wavelength coverage ($\mu$m)
& 2.5--5.0 & 4.6--9.2 & 7.2--13.4 & 17.5--26.5 
\\
Resolving power
& 120 & 53 & 50 & 48
\\
Exposure time (s)
& 400 & 196 & 245 & 442
\enddata
\end{deluxetable}

\clearpage
\begin{deluxetable}{l cc}
\tabletypesize{\scriptsize}
\tablewidth{0pt}
\tablecolumns{3}
\tablecaption{Slit positions of $AKARI$ spectroscopic observations
\label{tab_sppos}}
\tablehead {
\colhead{Slit position} &\colhead{NG/SG1/SG2} &\colhead{LG2}
}
\startdata 
Equatorial peak
&($\rm 11^h24^m29^s.1$, $\rm -59^\circ15'54''$)
&($\rm 11^h24^m29^s.9$, $\rm -59^\circ16'04''$)
\\
Ejecta peak
&($\rm 11^h24^m44^s.3$, $\rm -59^\circ16'40''$)
&($\rm 11^h24^m45^s.4$, $\rm -59^\circ16'51''$)
\\
Background
&($\rm 11^h25^m03^s.6$, $\rm -59^\circ16'24''$)
&($\rm 11^h25^m03^s.8$, $\rm -59^\circ16'40''$)
\\ 
\enddata
\tablecomments{
The slit position angles were fixed to be 44$^\circ$
by the satellite orbit.
}
\end{deluxetable}

\clearpage
\begin{deluxetable}{l ccc cc}
\tabletypesize{\scriptsize}
\tablewidth{0pt}
\tablecolumns{6}
\tablecaption{IR measurements of G292.0$+$1.8 \label{tab_f} }
\tablehead {
\colhead{Band} &\multicolumn{3}{c}{Whole remnant} &\multicolumn{2}{c}{IR-Ejecta} 
\\
\cline{2-6}
\colhead{} 
&\colhead{Total Flux}  
&\colhead{ER peak} &\colhead{OES peak}
&\colhead{Ejecta flux\tablenotemark{a}} 
&\colhead{Peak\tablenotemark{b}} 
\\
\colhead{} &\colhead{(Jy)} 
&\colhead{(MJy sr$^{-1}$)} &\colhead{(MJy sr$^{-1}$)}
&\colhead{(Jy)} &\colhead{(MJy sr$^{-1}$)}
}
\startdata
\underline{Measured flux/intensity} \\
S11                & 0.30 (0.02)   & 0.53 (0.26)  & 0.53 (0.24) 
& 0.013 (0.001)  & 0.37 (0.24) \\
L15                & 2.81 (0.14)   & 5.10 (0.36)  & 5.04 (0.36)
& 0.066 (0.009)  & 1.96 (0.26)  \\
L24                & 11.98 (0.72)  & 19.09 (1.37) & 18.94 (1.26)
& 0.072 (0.015)  & 2.60 (0.36) \\
N60                & 30.6 (6.1)    & 8.81 (2.53)  & 7.65 (1.83)
& 0.173 (0.035)  & 5.32 (1.47)  \\
Wide-S             & 30.0 (9.0)    & 6.76 (2.14) & 6.97 (2.20)
& 0.250 (0.075)  & 6.05 (1.95)  \\
Wide-L             & $<$ 6.8        & $<$ 6.50    & $<$ 6.58
& $<$ 0.048      & $<$ 6.87              \\
\underline{Color} \\
S11/L15            & 0.11 (0.01)     & 0.10 (0.05)  & 0.11 (0.05)
& 0.20 (0.03)    & 0.19 (0.13)      \\
L15/L24            & 0.23 (0.02)     & 0.27 (0.03)  & 0.27 (0.03)
& 0.92 (0.23)    & 0.75 (0.14)           \\
N60/Wide-S         & 1.02 (0.37)     & 1.30 (0.56)  & 1.10 (0.43)
& 0.69 (0.25)    & 0.88 (0.37)      \\
Wide-S/Wide-L      & $>$ 4.42       & $>$ 1.04   & $>$ 1.06
& $>$ 5.25       & $>$ 0.88
\enddata
\tablenotetext{a}{
The area used for the flux measurement is indicated 
in the inset of Figure~\ref{fig_cplot}. 
}
\tablenotetext{b}{Measured intensity at the peak of SG slit position.}
\end{deluxetable}

\clearpage
\begin{deluxetable}{lc cc}
\tabletypesize{\scriptsize}
\tablewidth{0pt}
\tablecolumns{4}
\tablecaption{Results of $AKARI$ IRC spectroscopy
\label{tab_line}}
\tablehead {
\colhead{Line} &\colhead{Wavelength}
&\colhead{ER peak} &\colhead{IR-Ejecta peak}
\\
\colhead{} &\colhead{($\mu$m)} 
&\colhead{(erg cm$^{-2}$ s$^{-1}$ sr$^{-1}$)}
&\colhead{(erg cm$^{-2}$ s$^{-1}$ sr$^{-1}$)}
}
\startdata
~[Ar~II]   & 7.0   & $< 1.6\times10^{-6}$ & $< 2.9\times10^{-6}$ \\
~[Ar~III]   & 9.0   & $< 4.0\times10^{-7}$ & $< 1.1\times10^{-6}$ \\
~[Ne~II]   &12.8   & $< 2.7\times10^{-6}$ & $2.8\times10^{-5}$ \\
\enddata
\tablecomments{
Undetected lines below 3$\sigma$ detection limit. 
}
\end{deluxetable}


\begin{thebibliography}{}
\bibitem[Allen \& Dupree(1969)]{all69} Allen, J. W., \& Dupree, A. K.
  1969, \apj, 155, 27
\bibitem[Blair et al.(2007)]{bla07} Blair, W. P., 
  Ghavamian, P., Long, K. S., Williams, B. J., Borkowski, K. J., 
  Reynolds, S. P., \& Sankrit, R.
  2007, \apj, 662, 998
\bibitem[Bouchet et al.(2004)]{bou04} Bouchet, P.,
  De Buizer, J. M., Suntzeff, N. B., Danziger, I. J., Hayward, T. L.,
  Telesco, C. M., \& Packham, C.
  2004, \apj, 611, 394
\bibitem[Braun et al.(1986)]{bra86} Braun, R.,
  Goss, W., M., Caswell, J. L., \& Roger, R. S.
  1986, \aap, 162, 259
\bibitem[Camilo et al.(2002)]{cam02} Camilo, F., 
  Manchester, R. N., Gaensler, B. M., Lorimer, D. R., \& Sarkissian, J.
  2002, \apjl, 567, L71
\bibitem[Chevalier(1982)]{che82} Chevalier, R. A.
  1982, \apj, 259, 302
\bibitem[Chevalier(2005)]{che05} Chevalier, R. A.
  2005, \apj, 619, 839
\bibitem[Couderc(1939)]{cou39} Couderc, P.
  1939, Ann. d'Astrophys., 2, 271
\bibitem[Douvion et al.(2001)]{dou01} Douvion, T.,
  Lagage, P. O., Cesarsky, C. J., Dwek, E.
  2001, \aap, 373, 281
\bibitem[Draine \& Lee(1984)]{dra84} Draine, B. T., 
  \& Lee, H. M.
  1984, \apj, 285, 89
\bibitem[Draine(2003)]{dra03} Draine, B. T.,
  2003, \apj, 598, 1017
\bibitem[Dwek et al.(1987)]{dwe87} Dwek, E.,
  Petre, R., Szymkowiak, A., \& Rice, W. L.
  1987, \apj, 320, 27
\bibitem[Dwek et al.(1996)]{dwe96} Dwek, E.,
  Foster, S. M., \& Vancura, O.
  1996, \apj, 457, 244
\bibitem[Dwek \& Arendt(2008)]{dwe08} Dwek, E. \& Arendt, R. G.
  2008, \apj, 685, 976
\bibitem[Edo(1983)]{edo83}
    Edo, O. 1983, PhD Dissertation, Dept. of Physics, University of Arizona
\bibitem[Fesen \& Gunderson(1996)]{fes96} Fesen, R. A. \& Gunderson, K. S.
  1996, \apj, 470, 967
\bibitem[Gaensler \& Wallace(2003)]{gae03} Gaensler, B. M. \&
  Wallace, B. J.
  2003, \apj, 594, 326
\bibitem[Gaustad et al.(2001)]{gau01} Gaustad, J. E., 
  McCullough, P. R., Rosing, W., \& Van Buren, D. 
  2001, PASP, 113, 1326
\bibitem[Ghavamian et al.(2005)]{gha05} Ghavamian, P.,
  Hughes, J, P. \& Williams, T. B.
  2005, \apj, 635, 365
\bibitem[Ghavamian et al.(2009)]{gha09} Ghavamian, P.,
  Raymond, J. C., Blair, W. P., Long, K. S., Tappe, A., 
  Park, S., Winkler, P. F.
  2009, \apj, in press
\bibitem[Glassgold at al.(2007)]{gla07} Glassgold, A. E.,
  Najita, J. R., \& Igea, J.
  2007, \apj, 656, 515
\bibitem[Gonzalez \& Safi-Harb(2003)]{gon03} Gonzalez, M.,
  \& Safi-Harb, S.
  2003, \apjl, 583, 91
\bibitem[Goss at al.(1979)]{gos79} Goss, W. M.,
  Shaver, P. A., Zealey, W. J., Murdin, P., \& Clark, D. H.
  1979, \mnras, 188, 357
\bibitem[Green et al.(2004)]{gre04} Green, D. A.,
  Tuffs, R. J., \& Popescu, C. C.
  2004, \mnras, 355, 1315
\bibitem[Hines et al.(2004)]{hin04} Hines, D. C., et al.
  2004, \apjs, 154, 290
\bibitem[Hughes et al.(2003)]{hug03} Hughes, J. P., et al.
  2003, \apjl, 591, L139
\bibitem[Hui \& Becker(2006)]{hb06} Hui, C. Y., \& Becker, W. 
  2006, \aap, L457, 33
\bibitem[Hwang et al.(2004)]{hwa04} Hwang, U., et al.
  2004, \apjl, 615, 117
\bibitem[Jones(2004)]{jon04} Jones, A. P. 
  2004 in ASP Conf. Ser. 309, Astrophysics of Dust,
  ed. A. N. Witt, G. C. Clayton, \& B. T. Draine (San Francisco: ASP), 347
\bibitem[Kawada et al.(2007)]{kaw07} Kawada, M., et al.
  2007, PASJ, 59, 389
\bibitem[Koo et al.(2007)]{koo07} Koo, B.-C.,
  Moon, D.-S., Lee, H.-G., Lee, J.-J., \& Matthews, K.
  2007, \apj, 657, 308
\bibitem[Krause et al.(2005)]{kra05} Krause, O., et al.
  2005, Science, 308, 1604
\bibitem[Laor \& Draine(1993)]{lao93} Laor, A., 
  \& Draine, B. T. 
  1993, \apj, 402, 441
\bibitem[Lockhart et al.(1977)]{loc77} Lockhart, I. A., 
  Gross, W. M., Caswell, J. L., \& McAdam, W. B.
  1977, \mnras, 179, 147 
\bibitem[Mathis et al.(1977)]{mat77} Mathis, J. S., 
  Rumpl, W., \& Nordsieck, K. H.
  1977, \apj, 217, 425
\bibitem[Matzner \& McKee(1999)]{mat99} Matzner, C. D., \& McKee, C. F.
  1999, \apj, 510, 379
\bibitem[Moon et al.(2004)]{met04} 
  Moon, D.-S., et al. 2004, \apjl, 610, L33
\bibitem[Moon et al.(2009)]{moo09} 
  Moon, D.-S., et al. 2009, \apjl, 703, L81
\bibitem[Murakami et al(2007)]{mur07} Murakami, H., et al.
  2007, PASJ, 59, 369
\bibitem[Murdin \& Clark(1979)]{mur79} Murdin, P. \& Clark, D. H.
  1979, \mnras, 189, 501
\bibitem[Ng \& Romani(2007)]{ng07} Ng, G.-Y. \& Romani, R. W.
  2007, \apj, 660, 1357
\bibitem[Nozawa et al.(2003)]{noz03} Nozawa, T.,
  Kozasa, T., Umeda, H., Maeda, K. \& Nomoto, K.
  2003, \apj, 598, 785
\bibitem[Nozawa et al.(2006)]{noz06} Nozawa, T.,
  Kozasa, T., \& Habe, A.
  2006, \apj, 648, 435
\bibitem[Ohyama et al.(2007)]{ohy07} Ohyama, Y., et al.
  2007, PASJ 59, 411
\bibitem[Onaka et al.(2007)]{ona07} Onaka, T., et al.
  2007, PASJ 59, 401
\bibitem[Park et al.(2002)]{par02} Park, S.,
  Roming, P. W. A., Hughes, J. P., Slane, P. O., Burrows, D. N.,
  Garmire, G. P., \& Nousek, J. A.
  2002, \apjl, 564, L39
\bibitem[Park et al.(2004)]{par04} Park, S., et al.
  2004, \apjl, 602, L33
\bibitem[Park et al.(2007)]{par07} Park, S.,
  Hughes, J. P., Slane, P. O., Burrows, D. N., Gaensler, B. M., 
  \& Ghavamian, P.
  2007, \apjl, 670, L121
\bibitem[Pittard et al.(2001)]{pit01} Pittard, J. M.,
  Dyson, J. E., Falle, S. A. E. G., \& Hartquist, T. W.
  2001, \aap, 375, 827
\bibitem[Rho et al.(2008)]{rho08} Rho, J.,
  Kozasa, T., Reach, W. T., Smith, J. D., Rudnick, L., DeLaney, T.,
  Ennis, J. A., Gomez, H., \& Tappe, A.
  2008, \apj, 673, 271
\bibitem[Rho et al.(2009)]{rho09} Rho, J.,
  Reach, W. T., Tappe, A., Hwang, U., Slavin, J. D., Kozasa, T., \& Dunne, L.,
  2009, \apj, 700, 579
\bibitem[Rodgers et al.(1960)]{rod60} Rodgers, A. W., 
  Campbell, C. T., \& Whiteoak, J. B. 
  1960, \mnras, 121, 103
\bibitem[Sakon et al.(2008)]{sak08} Sakon, I., et al. 
  2008, Proc. of SPIE, 7010, 88
\bibitem[Semenov et al.(2003)]{sem03}
    Semenov, D., Henning, Th., Helling, Ch., Ilgner, M., \& Sedlmayr, E.
    2003, \aap, 410, 611
\bibitem[Smith et al.(2009)]{smi08} Smith, J. D. T.,
  Rudnick, L., Delaney, T., Rho, J., Gomez, H., Kozasa, T.,
  Reach, W., \& Isensee, K.
  2009, \apj, 693, 713
\bibitem[Temin et al.(2006)]{tem06} Temim, T., et al.
  2006, \aj, 132, 1610
\bibitem[Tielens et al.(1994)]{tie94} Tielens, A. G. G. M., 
  McKee, C. F., Seab, C. G., \& Hollenbach, D. J.
  1994, \apj, 431, 321
\bibitem[Vancura et al.(1994)]{van94} Vancura, O., 
  Raymond, J. C., Dwek, E., Blair, W. P., Long, K. S., \& Foster, S. 
  1994, \apj, 431, 188
\bibitem[Wheeler et al.(2008)]{whe08} Wheeler, J. C.,
  Maund, J. R., \& Couch, S. M.
  2008, \apj, 677, 1091
\bibitem[Williams et al.(2006)]{wil06} Williams, B. J., et al. 
  2006, \apjl, 652, L33
\bibitem[Williams et al.(2008)]{wil08} Williams, B. J., et al. 
  2008, \apj, 687, 1054
\bibitem[Winkler \& Long(2006)]{win06} Winkler, P. F. \&
  Long, K. S.
  2006, \aj, 132, 360
\bibitem[Winkler \& Petre(2007)]{win07} Winkler, P. F., \& Petre, R.
  2007, \apj, 670, 635
\bibitem[Winkler et al.(2009)]{win08} Winkler, P. F.,
  Twelker, K., Reith, C. N., \& Long, K. S.
  2009, \apj, 692, 1489
\bibitem[Zeiger et al.(2008)]{zet08}  
  Zeiger, B. R., Brisken, W. F., Chatterjee, S., \& Goss, W. M.  
  2008,  \apj, 674, 271
\bibitem[Zubko et al.(2004)]{zub04}  Zubko, V.,
  Dwek, E., \& Arendt, R. G.
  2004, \apjs, 152, 211
\end{thebibliography}
\end{document}